\newcommand{\Msun}{\rm M_\odot}
\newcommand{\gammaad}{\hat{\gamma}}
\newcommand{\gammap}{{\gamma_p}}
\newcommand{\ein}{\tilde{E}_{\rm in}}
\newcommand{\Ein}{E_{\rm in}}
\newcommand{\Mej}{M_{\rm ej}}
\newcommand{\Ek}{E_k}
\newcommand{\rhoht}{\tilde{\rho}_h}
\newcommand{\ek}{\tilde{E}_k}
\newcommand{\mint}{\tilde{m}}
\newcommand{\Mx}{m_{\rm ex}}
\newcommand{\mx}{\tilde{m}_{\rm ex}}
\newcommand{\mxo}{\tilde{m}_{\rm ex}^\prime}
\newcommand{\alphanr}{\alpha_{\rm nr}}
\newcommand{\alphar}{\alpha_{\rm r}}
\newcommand{\alphag}{\alpha_{\rm g}}

\documentclass[preprint2]{aastex}
\begin{document}
\title{Trans-Relativistic Blast Waves in Supernovae as \\Gamma-Ray Burst Progenitors} 
\author{Jonathan C. Tan\altaffilmark{1}, Christopher D. Matzner\altaffilmark{2}, and Christopher F. McKee\altaffilmark{1,3}}
\affil{1. Department of Astronomy, University of California, Berkeley, CA 94720, USA.}
\affil{2. CITA, University of Toronto, 60 St. George Street Toronto, Ontario M5S 3H8, Canada.}
\affil{3. Department of Physics, University of California, Berkeley, CA 94720, USA.}

\begin{abstract}
We investigate the acceleration of shock waves to relativistic
velocities in the outer layers of exploding stars. By concentrating
the energy of the explosion in the outermost ejecta, such
trans-relativistic blast waves can serve as the progenitors of
gamma-ray bursts (GRBs); in particular, the ``baryon-loading'' problem
that plagues many models of GRBs is
circumvented.  Postshock acceleration is effective in boosting the
kinetic energy in relativistic ejecta. 
We present physically motivated analytic expressions to describe 
trans-relativistic blast waves in supernovae, and we validate these
expressions against numerical simulations of test problems.
Investigating the effect of stellar structure on mass ejection, we
find that relativistic ejecta are enhanced in more centrally condensed
envelopes---e.g., for radiative envelopes, when the luminosity
approaches the Eddington limit. Convenient formulae are presented with
which to estimate the production of relativistic ejecta from a given
progenitor.

We apply our analytic and numerical methods to a model of SN 1998bw,
finding significantly enhanced relativistic ejecta compared to
previous studies. We propose that GRB 980425 is associated with SN
1998bw and may have resulted from an approximately spherical explosion
producing $\sim10^{-6}\:{\rm M_{\odot}}$ of mildly relativistic ejecta
with mean Lorentz factor $\bar{\Gamma}\sim 2$, which then interacted
with a dense circumstellar wind with mass loss rate $\sim{\rm
few}\times 10^{-4}\:{\rm M_{\odot}\:yr^{-1}}$. A highly asymmetric
explosion is not required.  An extreme model of ``hypernova''
explosions in massive stars is able to account for the energetics and
relativistic ejecta velocities required by many of the observed
cosmological GRBs. However, the most energetic bursts require
asymmetric expulsion of ejecta, perhaps caused by rotationally
flattened progenitors. We present simplified models and simulations of
explosions resulting from accretion-induced collapse of white dwarfs
and phase transitions of neutron stars.  While we find increased
energies in relativistic ejecta compared to previous studies, these
explosions are unlikely to be observed at cosmological distances with
current detectors, unless extreme explosion energies and asymmetries
are invoked.
\end{abstract}


\section{Introduction}\label{S:Intro}
More than 30 years after their discovery, gamma-ray bursts (GRBs)
remain one of the most outstanding puzzles in astrophysics. A
``relativistic fireball'' model has emerged to explain the
$\gamma$-ray emission and successfully predict the existence of longer
wavelength afterglows. However, the identity, or identities, of the
central engine driving such fireballs is literally shrouded in
uncertainty. At least two major problems stalk the proposed GRB
models. Firstly, extreme isotropic energies are implied by the
cosmological distances of some bursts, determined from redshifted
absorption lines in their afterglows. Secondly, the ``baryon loading -
compactness problem'' results from the large energies, small sizes
(implied by short timescale variability) and optically thin spectra of
GRBs and requires the source to move towards us highly
relativistically. This in turn mandates miniscule mass-loading, $\sim
5.6\times 10^{-5} E_{\gamma,52}(\bar{\Gamma}/100)^{-1} \epsilon_\gamma^{-1}
\:{\rm M_{\odot}}$, where $E_{\gamma,52}$ is the
isotropic $\gamma$-ray energy in units of $10^{52}\:{\rm ergs}$,
$\bar{\Gamma}\equiv (1-\bar{\beta}^2)^{-1/2}$ is the mean Lorentz
factor of the source mass ($\bar{\beta}$ is the mean velocity relative
to the speed of light) and $\epsilon_\gamma$ is the conversion efficiency
of bulk kinetic energy to $\gamma$-rays \citep[see, e.g.,][for a review]{pir99}. 

One mechanism that can naturally overcome these hurdles is the
acceleration of shocks in a progenitor, which confines and efficiently
taps the energy from a central explosion resulting from the formation
of a compact stellar mass object.  Although the shock is initially
non- or mildly relativistic, it may accelerate to relativistic
velocities and channel a fraction of the total explosion energy into
relativistic ejecta. Shock acceleration\footnote{We shall use the
term ``shock acceleration'' to refer to this hydrodynamical effect,
which ought not be confused with particle acceleration in shocks.}
occurs in the presence of a steeply declining density gradient, and so
only a small fraction of the ejecta mass reaches relativistic speeds.

This mechanism has provided the basis for many different models of
GRBs, ranging from normal supernovae \citep[in the first astrophysical
GRB model;][]{col74}, to extremely energetic supernovae or ``hypernovae''
\citep{pac98}, to phase transitions of neutron stars into strange
stars \citep{fry98}. Accretion-induced collapse of white dwarfs to
neutron stars \citep{woo92,fry99} may also drive a shock into the
outer atmosphere of the white dwarf, but such shocks have so far been
examined only in the context of much weaker Type Ia supernova
explosions \citep{ber96}. Shock acceleration is also an important
factor in aspherical models such as those involving
rotationally flattened progenitors \citep{che89,1999ApJ...524L.107K}
and jet-like explosions of the core \citep{mac99,hof99}. Even if a jet
pierces the stellar envelope entirely \citep[as in the model
of][]{alo00}, the production of relativistic ejecta by shock
acceleration may dominate the observed GRB in most
directions.

While the propagation of shock waves is well understood in the
nonrelativistic \citep{sed46, sed59, tay50, ost88, mat99} and
ultrarelativistic \citep{joh71,bla76} regimes, similarity solutions do
not exist for the mildly relativistic case. Thus, previous analytic
models of the acceleration of shocks at these intermediate velocities
have extrapolated nonrelativistic results with simple scaling laws
\citep[e.g.][]{gna85} intermediate between the non- and
ultrarelativistic limits. Furthermore, the postshock evolution of
trans-relativistic ejecta has received only a crude treatment, if any,
in previous work.

The aim of this paper is to provide a quantitative, theoretical
description of trans-relativistic shock and postshock acceleration,
which will be of use in many GRB models. A {\it trans-relativistic
blast wave} is one in which some of the mass is accelerated to highly
relativistic velocities, but most of the mass is not; in particular
the ratio of the explosion energy to the rest energy of the ejecta is
$< {\cal O}(1)$. In \S\ref{S:theory}, utilizing a relativistic
one-dimensional Lagrangian hydrodynamic code based on the method of
exact Riemann solvers \citep{mar94}, we investigate the dynamics of
shocks and shocked ejecta in idealized model density distributions. We
present physically motivated analytic approximations to describe the
shock and postshock acceleration as well as the kinetic energy
distribution of ejecta from density distributions approximating
polytropes at their edges.
We explore how a progenitor's structure affects its ejecta in
\S\ref{S:whips}, finding that highly condensed envelopes produce fast or
relativistic ejecta most efficiently and that this corresponds to
stars whose luminosities approach the Eddington limit. We present
formulae to estimate the energy in relativistic ejecta from the
parameters of a given progenitor star.
Our results, although primarily derived for spherically symmetric
explosions, can also be applied to sectors of aspherical explosions, a
topic we address in \S \ref{S:aspherical}. Explosions where the effect
of gravity is strong, for example relevant to ejecta from neutron
stars, are treated in \S \ref{S:WD/NS-Theory}.

We apply our results to four astrophysical examples, as
potential GRB engines (\S\ref{S:applications}). First, and in most
detail, we consider supernova explosions in the exposed cores of
massive post main-sequence stars (\S\ref{S:98bw}). The probable
association of supernova (SN) 1998bw with GRB 980425 \citep{gal98} has
sparked renewed interest in such models. Other supernovae for which a
suspicion of GRB association exists include SN 1997cy with GRB 970514
\citep{ger00, 2000ApJ...534L..57T} and SN 1999E with GRB 980901 
\citep{tho99}. The late time excess flux, relative to that expected 
from a power-law decline, of the afterglows of GRB 970228
\citep{fru99} and GRB 980326 \citep{blo99} has been cited as evidence 
for the presence of supernovae \citep{blo99, rei99, gal00a}. However,
alternative mechanisms of afterglow re-radiation \citep{wax00} and
scattering \citep{esi00} by dust, have also been proposed to produce
the excess flux. Cosmological GRB afterglows are associated with the
stellar population of their host galaxies \citep{blo00b}, but there is
only tentative evidence for a direct association with star forming
regions \citep{kul00,djo00,gal00b}, which would be expected if they
arose from the deaths of massive stars.

If there is an association between GRB 980425 and SN 1998bw, then the
$\gamma$-ray energy is only $\sim 10^{48}\:{\rm ergs}$, thousands to
millions of times weaker than the inferred isotropic energies of the
cosmological bursts; moreover, its associated galaxy is much too near
to be drawn from the same distribution as the other known host
galaxies \citep{1999ApJ...520...54H}. This event may be an example of
a subclass of GRBs, termed S-GRBs \citep{blo98}. \citet{woo99}
simulated an energetic ($\sim {\rm few}\times 10^{52}\:{\rm ergs}$)
explosion in a variety of carbon-oxygen and helium star progenitors,
attempting to fit the observed supernova lightcurve. They considered
the possibilities of $\gamma$-ray emission from shock break-out and
the interaction of ejecta with a circumstellar wind. With
nonrelativistic codes they found the simple spherical case unable to
produce enough relativistic ejecta to account for even this weak
GRB. This result, together with the extremely large isotropic energies
($E_{\gamma}\sim 10^{51} - 10^{54}\:{\rm ergs}$) of the cosmological
bursts, has focussed attention on aspherical explosions involving
jets, which may be able to explain the different types of GRBs as a
function of the viewing angle to the jet
\citep{whe00a, whe00b, mac99, mac00, alo00, wan00}. However,
uncertainties in previous models of trans-relativistic dynamics
warrant further investigation of the simplest spherical case. 
Indeed, our relativistic analysis of the model considered by
\citet{woo99} yields sufficient mildly relativistic ejecta to account
for GRB 980425 and radio observations of SN 1998bw, for a progenitor
embedded in a dense wind with mass loss rate $\sim{\rm few}\times 10^{-4}\:{\rm
M_{\odot}\:yr^{-1}}$.
Our analysis therefore validates the suggestion by \cite{mat99} that
GRB 980425 could be the product of a {\em spherical} explosion of SN
1998bw. 

In \S\ref{S:C-GRBs} we continue with explosions in the cores of
massive stars, but now at much greater energies. These
hypernovae\footnote{Unlike \cite{iwa98}, we shall reserve the term
``hypernovae'' for much more energetic events than SN 1998bw.} are
hypothesized to result from magnetic extraction of the rotational
energy of nascent neutron stars or black holes \citep{pac98}. We
consider an extreme example with $5\times 10^{54}\:{\rm ergs}$, which
might arise from the formation of a $\sim 10\:{\rm M_{\odot}}$ black
hole. For the purposes of this crude estimate, we embed the explosion
in the $\sim 5\:{\rm M_\odot}$ envelope ejected from our SN 1998bw
progenitor, and calculate the resulting energy distribution. To
compare to the observed cosmological bursts we derive conservative
constraints on the minimum mean Lorentz factor of ejecta carrying the
$\gamma$-ray energy, by considering the timescale of the burst, and
requiring the implied circumstellar material to be optically thin to
the burst photons.  We also consider the minimum Lorentz factor
required for the optical depth due to photon-photon pair production to
be small.  We find that the overall energetics and relativistic ejecta
velocities required by many cosmological bursts can be explained by
such a simple, though admittedly extreme, model. However there are
some events, notably GRB 990123, that lie well beyond the reach of the
spherical model. Considering aspherical explosions (\S
\ref{S:aspherical}), we calculate that enhancements in shock strength
along the preferred axis, perhaps set by rotation, by factors of only
a few are required to explain all cosmological GRBs.  Interaction of
the relativistic ejecta with inhomogeneities may give bursts some
short timescale substructure \citep{der99,fen99}, though this can only
account for a small fraction of the total burst energy \citep{sar97}.
Instabilities associated with an accelerating shock
\citep{1994ApJ...435..815L} may create postshock velocity
perturbations that also lead to GRB variability.

Finally we consider simple models of explosions in white dwarfs and
neutron stars (\S\ref{S:WDNS}), in order to investigate shock and
postshock acceleration in their outer layers. Such shocks may be
driven by the accretion-induced collapse of a white dwarf to a neutron
star or the phase transition of a neutron star to a hypothesized
``strange'' star, composed of more stable strange quark matter.  For
the production of relativistic ejecta, these models have the advantage
that large explosion energies may be coupled to very small ejecta
masses. Given a total ejecta mass and energy, as predicted from
detailed core-collapse models \citep{fry98,fry99}, we estimate the
energy distribution of the ejecta. Again, we find enhanced amounts of
energy in relativistic ejecta compared to previous estimates. We
compare to cosmological GRBs, this time by combining the burst
timescale with an assumed density of a uniform ambient medium. We find
that white dwarf explosions are unlikely to be relevant to these
bursts. Our neutron star models, while more energetic, need to be
pushed to extreme limits to provide observable fluxes at cosmological
distances.

\section{Theory}\label{S:theory}
The process of mass ejection in a supernova explosion proceeds in a
sequence of distinct dynamical phases \citep{mat99}. In the first
phase, the stellar envelope is engulfed by the outward-propagating
blast wave launched by collapse of the core. Except perhaps in the
vicinity of the central engine and within the shock itself, the blast
wave is adiabatic. For an explosion in the tenuous outer envelope of a
supernova, the postshock pressure is dominated by photons; for a
relativistic shock in a compact object, it may be dominated by
relativistic motions of leptons \citep{wea76}. In either case, the
ratio of specific heats is $\gammaad= 4/3$. The boundary of this blast
wave is a shock front, which imprints its velocity (and an associated
specific entropy) on the postshock gas. The shock velocity decreases
in most parts of the star, but it increases in regions with a
sufficiently steep density profile -- an effect analogous to the
cracking of a whip. As the postshock pressure is dominated by
radiation pressure, the shock front has a finite optical depth (for
which photons diffuse upstream as fast as gas flows downstream). Thus
the shock breaks out of the star at some finite depth. Only for a
sufficiently energetic explosion in a sufficiently compact star will
the shock accelerate to relativistic velocities before it breaks
out. In \S\ref{S:shockprop} we investigate shock acceleration in such
progenitors.

After the stellar envelope has been shocked, its thermal and kinetic
energies are approximately equal. The next phase of evolution is one
of postshock acceleration, in which heat is converted into outward
motion and the ejecta approach a state of free expansion. For the gas
in the outer layers of the star, this increases the velocity by over a
factor of two compared to the postshock state if the shock is
nonrelativistic \citep{mat99}. However, gas hit by a relativistic
shock undergoes even more dramatic acceleration, because postshock
kinetic and thermal energy densities are both dominated by photons,
which eventually transfer their energy to the baryons in the freely
expanding state. This conversion of internal energy to kinetic leads
to a large increase in the Lorentz factor during the acceleration
phase. The acceleration is accentuated by the fact that each gas
element is pushed outward by interior ejecta, which though slower, are
at higher pressure (\S\ref{shacc}).

Each stage of this process has been the study of previous theoretical
investigations; our goal is to investigate in greater detail the
connection between the shock velocity and the final expansion
velocity, across the transition from Newtonian to relativistic
motion. To describe this process analytically, we present formulae that
have been calibrated against numerical simulations for a variety of
progenitor stars (\S\S \ref{S:shockprop} and \ref{shacc}). In
\S\ref{S:KEdist} we give an analytical prediction of the
distribution of energy with Lorentz factor in the state of free
expansion, i.e., the information necessary to calculate the
circumstellar interaction and the resulting burst of energetic
radiation. 
We consider the dependence of relativistic mass ejection on the
progenitor's structure in \S\ref{S:whips}. We apply our dynamical
results to sectors of aspherical explosions in \S\ref{S:aspherical}
and explosions where the ejecta are strongly affected by gravity in
\S\ref{S:WD/NS-Theory}.  

\subsection{Shock Propagation}\label{S:shockprop} 

\begin{figure}
\epsscale{1.0} \plotone{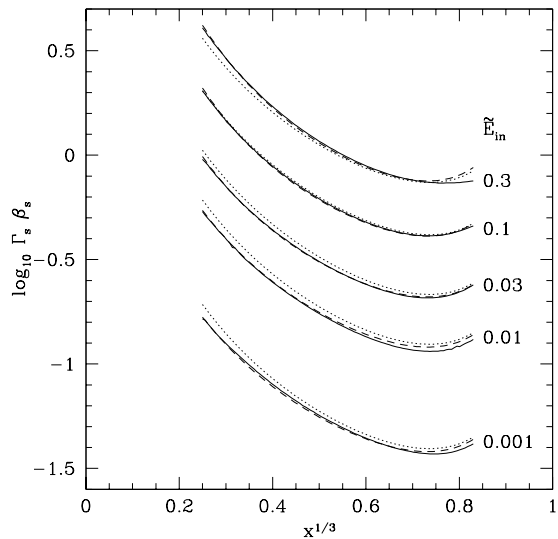}
\caption{Spherical shock propagation down a density profile 
$\rho=\rho_h x^3$, where $x\equiv 1-r/R$, for different explosion
energies, $\ein$. Simulation results ({\it solid} lines) are compared
to the analytic prediction of this paper (eqs. [\ref{relshockf}] and
[\ref{relshock}]) ({\it dashed} lines), and to the variation of the
scaling predicted by \citet{gna85} (GMM scaling) as expressed by
equation (\ref{GMM}) ({\it dotted} lines). The normalization constant,
$A$, is 0.68 for $\ein\leq0.03$, and 0.72, 0.74 for $\ein=0.1,0.3$,
respectively.
\label{fig:propagation}}
\end{figure}

\begin{figure}
\epsscale{1.0} \plotone{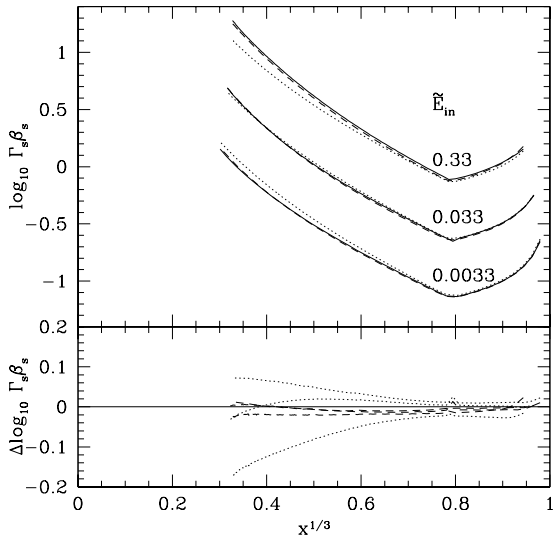}
\caption{Spherical shock propagation down a realistic stellar density
profile, given by equation (\ref{rhostellar}). The differences between
the GMM scaling (eq. [\ref{GMM}]) ({\it dotted} lines) and our
improved formula (eqs. [\ref{relshockf}] and [\ref{relshock}]) ({\it
dashed} lines) are clearly revealed in the outer region
($x^{1/3}<0.8$), where the density gradient is very steep. The
simulation results are again shown by the {\it solid} lines. The
normalization constant, $A$, is 0.736 for all energies. The lower
panel shows the ratio of the analytic predictions to the numerical
results. For both cases $\ein=0.0033,0.03,0.3$ from top to bottom. The
GMM scaling under predicts relativistic shock acceleration and
over-predicts nonrelativistic. The final Lorentz factor of ejecta
depend sensitively on $\Gamma_s$ (\S\ref{shacc}), and the observed
difference, $\sim 33\%$, between the two laws results in a difference
of a factor of $\sim3$ in final Lorentz factors.
\label{fig:propagation2}}
\end{figure}

The propagation of a nonrelativistic blast wave through the stellar
interior and the subsequent acceleration phase were studied in detail
by \citet{mat99}. The shock velocity, $\beta_s c$, responds to two
competing trends. First is the overall deceleration as mass is
entrained through the shock, $\beta_s \propto
[\ein/\mint(r)]^{1/2}$, where 
\begin{equation}\label{mtildedefn}
\mint(r) \equiv \frac{m(r) - M_{\rm rem}}{\Mej}; ~~~~~ 0\leq \mint\leq 1
\end{equation}
is the fraction of the total ejecta mass, $M_{\rm ej}$, within $r$
(excluding any remnant mass $M_{\rm rem}$),
and 
\begin{equation}\label{eindefn} 
\ein\equiv \frac{\Ein}{M_{\rm ej} c^2}
\end{equation} 
is the injected energy in units of the rest energy, which is $<{\cal
O}(1)$ for trans-relativistic explosions. Second is the acceleration
of the shock front down the declining density gradient, for which
$\beta_s \propto [\mint(r)\Mej /(\rho r^3)]^{\alphanr}$ where $\alphanr
\simeq 0.19$ for typical outer density distributions
\citep{sak60}. The shock propagation is a combination of these two trends, and it is
not surprising that a product of the two formulae yields an excellent
approximation to the shock velocity \citep{mat99}:
\begin{equation}
\label{nonrelshock}
\beta_s=A \left(\frac{\ein}{\mint} \right)^{1/2}
\left(\frac{\mint \Mej}{\rho r^3} \right)^{\alphanr}.
\end{equation}

The coefficient $A$ can be evaluated for a spherical Sedov blast wave
in the distribution $\rho \propto r^{-k_{\rho}}$, in which it takes
the value
\begin{equation}
\label{Anr}
A=\sigma^{-1/2}\left(\frac{4\pi}{3-k_{\rho}}\right)^{-\alphanr},
\end{equation}
where $\sigma\equiv \ein/(\mint\beta_s^2)$, constant for self-similar
blast waves, measures the fraction of the explosion energy that is
kinetic. Note that the shock accelerates in this region only if
$k_\rho>3$. \cite{mat99} found that the Primakoff distribution $k_\rho
= 17/7$ (for which $A = 0.794$) adequately describes the interiors of
red and blue supergiant progenitors of type II supernovae. However, we
have found that the accuracy of equation (\ref{nonrelshock}) is
enhanced if $A$ is evaluated for the typical value of $k_\rho$ in the
interior of progenitor. To make this evaluation, we use the pressure
gradient approximation of
\citet{ost88}, with $\gammaad = 4/3$:
\begin{equation}\label{sigma}
\sigma \simeq \frac{2(29-7k_\rho)(3-k_\rho)}{113-35k_\rho},
\end{equation}
which is exact for the Primakoff blast wave, $k_\rho = 17/7$.  For the
progenitor model CO6 of \citet{woo99}, which we use to simulate SN
1998bw (\S \ref{S:98bw}), we estimate $k_\rho\simeq 1.9$ which gives
$A\simeq 0.736$.

We also consider the velocity normalization for shocks that develop in
centrally concentrated, non-power-law distributions, with monotonically
steepening density gradients as $r\rightarrow R$. These types of
distribution mimic those of white dwarfs and neutron stars.  We
estimate a mean $k_\rho$ in the central region, where the shock
structure forms, by evaluating a moment of the whole density
distribution, $J_l\equiv\int_0^R(r/R)^l d(m/M)$. For a power law
distribution
\begin{equation}
k_\rho = \frac{(3+l)J_l-3}{J_l-1}.
\label{krhoJl}
\end{equation}
From numerical simulation (below), we find $l=-1$ gives an estimate of
an effective $k_\rho$ which accurately predicts $A$, via equations
(\ref{Anr}) and (\ref{sigma}), for $\ein \ll 1$, in several different
distributions we have tested. We find $A$ may have a weak energy
dependence for $\ein\sim{\cal O}(1)$ (below).

Although equation (\ref{nonrelshock}) is quite accurate in the
nonrelativistic regime, it breaks down when $\beta_s \rightarrow 1$
and the shock becomes trans-relativistic. Even in the context of a
planar, power-law density distribution, the dynamics are not
self-similar in this stage. Therefore, it has received little
attention in previous work, whereas the opposite limit of
self-similar, ultrarelativistic shock propagation was treated
approximately by \cite{joh71}.  However, a knowledge of the
nonrelativistic and ultrarelativistic shock velocity scalings does not
suffice to specify the relative coefficient connecting the two
stages. Moreover, it does not allow one to describe ejecta that are
only mildly relativistic; as we shall show, these trans-relativistic
ejecta contain more energy than the ultrarelativistic. For
these reasons, we wish to investigate in detail the transition between
Newtonian and ultrarelativistic motion.

Using an approximate treatment, \citet{joh71} presented the following
estimate of the shock Lorentz factor, $\Gamma_s \equiv
(1-\beta_s^2)^{-1/2}$, valid in the ultrarelativistic ($\Gamma_s\gg1$)
limit:
\begin{equation}
\label{ultrarelshock}
\Gamma_s \simeq \frac{1}{\sqrt{2}}\left(\frac{E_i}{\rho c^2}\right)^{\alphar},
\end{equation}
where $\alphar=\frac{1}{2}\sqrt{3}(2+\sqrt{3})^{-1}\simeq 0.232$, and
$E_i$ is the energy density at some reference point in the
flow. Johnson \& McKee obtained this result by assuming that initially
a region of hot gas, with a spatially uniform value of $E_i$, abutted
cold gas with decreasing density $\rho(r)$. As a result, the Riemann
invariant was the same on all forward characteristics emanating from
the hot gas. In the case considered here there is no region with
constant $E_i$, so the exact values of $\alpha_{r}$ and the numerical
coefficient may be somewhat different.

Noting the similarity between the Newtonian and relativistic indices
$\alphar$ and $\alphanr$ that relate the shock velocity to the
density profile, \citet{gna85} suggested the following formula for
shocks bounding spherical blast waves: 
\begin{equation}\label{gnatyk}
\Gamma_s \beta_s \propto (\rho r^{3})^{-\alpha_{\rm g}},
\end{equation}
where $\alpha_{\rm g} \simeq 0.5$ for decelerating and $0.2$ for
accelerating shocks. This approximation was found to be inferior to
equation~(\ref{nonrelshock}) in the nonrelativistic limit by
\cite{mat99}, but it does suggest that the quantity $\Gamma_s \beta_s$
can be approximated analytically. 

However, equation (\ref{gnatyk}) does not specify the normalization of
the shock velocity. We therefore consider instead the following
formula:
\begin{equation}\label{GMM}
(\Gamma_s \beta_s)_{\rm GMM} = A \left(\frac{\ein}{\mint} \right)^{1/2}
\left(\frac{\mint \Mej}{\rho r^3} \right)^{\alphag}. 
\end{equation}
In the Newtonian limit this formula reduces to a shock velocity
approximation very similar to equation (\ref{nonrelshock}), and it
makes a transition to relativistic motion in the manner suggested by
\citet{gna85}, (equation [\ref{gnatyk}]): hence the subscript GMM. Since
$\mint\simeq 1$ in the region of strong acceleration at the outside of
a star, this formula has the same scaling as equation (\ref{gnatyk})
there. Whereas \cite{woo99} chose the shock-velocity coefficient
numerically, equation (\ref{GMM}) uses our earlier results and those
of \cite{mat99} to prescribe it analytically. Note that it is Gnatyk's
exponent $\alphag = 0.2$, rather than $\alphanr=0.19$, that enters
this expression.  Insofar as equation (\ref{nonrelshock}) accurately
predicts the behavior of a nonrelativistic shock front with
$\alphanr=0.2$ (and we find that it does), equation (\ref{GMM})
represents the result of an extrapolation to relativistic velocities
using Gnatyk's scaling. It remains to be determined whether the
transition to relativistic motion is captured accurately by this
formula.

To study the dynamics of trans-relativistic blast waves, we have
developed a relativistic, one-dimensional Lagrangian hydrodynamics
code based on an exact Riemann solver \citep{mar94}. The Lagrangian
nature of our code is a key requirement for handling large ranges in
density with fine zoning, of particular relevance for simulating
shocks accelerating in stellar atmospheres. The code accurately
handles standard test problems, including Centrella's
\citeyearpar{cen86} proposed wall shock and Riemann shock tube
tests. Additionally the code recovers the ultrarelativistic
Blandford-McKee \citeyearpar{bla76} solution for spherical blast waves.

With this code we have conducted a suite of numerical calculations.
For much of this study we employ a ``stellar'' model density
distribution: an outer stellar envelope of polytropic index $n$ has
the density profile 
\begin{equation}\label{eq:outerpolytrope}
\rho =  \rho_h (R/r-1)^n, 
\end{equation}
in any region where the external mass $\Mx$ can be ignored compared
to $M_\star$ and where $n$ does not vary. If this region extends
inward to half the radius \citep[e.g., for polytropes with $n\geq
3$;][]{cha39} then the coefficient $\rho_h$ is equal to the density
there: $\rho_h=
\rho(r=R/2)$. For our stellar model, we join this outer profile to an
inner, power-law density distribution:
\begin{equation}\label{rhostellar} 
\rho = \rho_h \times 	\left\{ 
\begin{array}{lc}
\left(\frac{R}{r_c}-1\right)^n \left(\frac{r}{r_c}\right)^{-k_\rho}, & r < r_c \\ 
\left(\frac{R}{r}-1\right)^n, & r_c <r<R. 
\end{array}
\right.
\end{equation}
We choose a core radius $r_c \simeq R/2$, polytropic index $n=3.85$, and
power law index $k_\rho = 1.9$ to resemble the progenitor model for
SN~1998bw considered in \S~\ref{S:98bw}. In order to find analytical
formulae valid for various density distributions, we have additionally
considered an ``external power law envelope'' model,
\begin{equation}\label{extpow}
\rho = \rho_h x^n,
\end{equation}
also with $n=3$, where
\begin{equation}
x\equiv 1-\frac{r}{R}.
\end{equation}
Equation (\ref{extpow}) is the $x\rightarrow 0$ limit of the
polytropic profile (\ref{eq:outerpolytrope}), but significantly
underestimates $\rho$ in the interior (e.g., by a factor of $2^n$ at $x=1/2$).

After an injection of thermal energy in the central few zones, a shock
develops and propagates outwards, decelerating in the inner region,
where the density gradient is shallow, and accelerating in the outer
layers, where the density gradient is steep. We have investigated
explosions with $\ein$ ranging from $0.001$ to $0.33$. In each case,
the resolution was increased either until the numerical results
converged, or until a correction for incomplete convergence could be
ascertained.

Empirically, we find that the scaling proposed by \citet{gna85}, as
realized in equation (\ref{GMM}), is roughly correct but not
sufficiently accurate for a study of the energetics of relativistic
ejecta. In order to capture the change in the value of the exponent
from $\alphanr$ to $\alpha_r$, we adopt instead the following
generalization of equation (\ref{GMM}):
\begin{eqnarray} 
\Gamma_s \beta_s & = & p(1+p^2)^{0.12},\label{relshockf}\\ p & \equiv
& A \left(\frac{\ein}{\mint}\right)^{1/2}\left(\frac{\mint \Mej}{\rho
r^3}\right)^{\alphanr}.\label{relshock}
\end{eqnarray}
For the density distributions described above, our numerical results
give $\alphanr=0.187$, in agreement with the nonrelativistic
self-similar theory and equation (\ref{nonrelshock}). We adopt this
numerical value in our subsequent formulae. $\alphanr^{-1}$ enters in
many of our expressions and we set this equal to 5.35. In the
ultrarelativistic regime ($p\gg1$), equation (\ref{relshockf}) was
chosen to reproduce the approximate result $\alphar\simeq 0.232$
obtained by \citet{joh71}.

In Figures \ref{fig:propagation} and \ref{fig:propagation2} we compare
equations (\ref{GMM}) and (\ref{relshock}) to the results of our model
simulations of explosions in the ``external power law'' and
``stellar'' density distributions. The two shock velocity formulae
under consideration are extremely similar, differing by only very
small powers of their parameters in both the non- and
ultrarelativistic limits. Their differences are most clearly seen in
the stellar model, where density, and hence shock velocity, spans a
greater range.

Limited numerical resolution does not allow us to follow the
acceleration of a nonrelativistic shock into the ultrarelativistic
regime. Instead we have increased the explosion energy to investigate
acceleration in different velocity regimes. Our results show that our
improved shock propagation model (eq. [\ref{relshockf}]) is accurate to
within $\pm 5\%$ for all energies and density distributions
tested.\footnote{This is after allowing for the normalization
constant, $A$, to increase slightly with $\ein$ for energetic
explosions, e.g., for the external power law envelope model, the low
energy ($\ein\leq0.03$) explosions have $A=0.68$ as predicted from
eq. (\ref{krhoJl}), while for $\ein=0.1,0.3$, $A=0.72, 0.74$.}
Furthermore we find the simple Gnatyk scaling under predicts the
acceleration of relativistic shocks and over-predicts the acceleration
of nonrelativistic shocks. Our highest energy explosion of the stellar
model shows that the Gnatyk scaling underestimates, by $\sim 33\%$,
the value of $\Gamma_s$ reached by a shock that accelerates from
$\Gamma_s
\beta_s\sim 1$ to $\Gamma_s \beta_s\sim 10$. The final Lorentz factor,
$\Gamma_f$, achieved by shocked ejecta depends sensitively on
$\Gamma_s$ (see \S\ref{shacc}).  For example, postshock acceleration
amplifies the above decrement into a factor of three error in $\Gamma_f$.

We must note that our approximation (eqs. [\ref{relshockf}] \&
[\ref{relshock}]) gives $\Gamma_s
\propto (\ein/\mint)^{0.56}$ rather than the correct scaling
$\Gamma_s \propto (\ein/\mint)^{1/2}$ that obtains in spherical,
ultrarelativistic blast waves \citep{bla76}. Because of this, the
accuracy of our formula breaks down for blast waves that start out
ultrarelativistically; i.e. with $\ein>1$. We begin to see these
deviations by the time $\ein\sim0.3$, as the value of $A$ tends to
increase slightly ($\lesssim 10\%$), for a given density distribution.
Equations (\ref{relshockf}) and (\ref{relshock}) are strictly valid
only for trans-relativistic blast waves, i.e. for shocks that are non-
or semi-relativistic to begin with, but that may accelerate to high
$\Gamma_s$, i.e. $\ein\lesssim {\cal O}(1)$. This is well satisfied
for most of the astrophysical contexts to which we shall apply the
formula in \S\ref{S:applications}, e.g., our model for supernova
1998bw has $\ein\simeq 0.003$. However, the extreme hypernova models
considered in \S\ref{S:C-GRBs} are at the limits of this
approximation's validity, with $\ein\simeq0.6$.

\subsection{Postshock Acceleration}\label{shacc}

After the entire stellar envelope has been traversed by the shock
wave, it is left in a state of outward motion. However, its heat
content is comparable to its kinetic energy, so it will accelerate
considerably as heat is converted into motion. Assuming a strict
conservation of lab-frame energy in each mass element -- from the
postshock state to a state of free expansion -- allows one to roughly
estimate the magnitude of this acceleration. However, this is a
significant underestimate of the acceleration since the outermost
layers gain energy at the expense of the slower ejecta within them,
because of a steeply declining pressure gradient. 

The relativistic strong shock jump conditions give \citep{bla76}
\begin{equation}
\label{jump}
\Gamma_s^2 = \frac{(\Gamma_2 +1)[\gammaad(\Gamma_2 -1)+1]^2}{\gammaad (2-\gammaad)(\Gamma_2 -1) +2},
\end{equation}
where the subscript $2$ refers to the fluid immediately
behind the shock. For $\gammaad=4/3$, we have $\beta_s/\beta_2=7/6$ and
$\Gamma_s/\Gamma_2=\sqrt{2}$ in the non- and ultrarelativistic limits,
respectively.

Now consider the postshock acceleration of material that was struck by
a nonrelativistic, strong shock as it approached the periphery of the
progenitor star. Thermal and kinetic energies are in perfect
equipartition behind the shock. Therefore, strict energy conservation
in each element implies $\beta_f/\beta_2=\sqrt{2}$ and so
$\beta_f/\beta_s = 6\sqrt{2}/7=1.21$. This should be contrasted with
the results of self-similar shock acceleration and postshock
acceleration, which for $\gammaad = 4/3$ yield $\beta_f/\beta_s = 2.16$
and $2.04$ for $n=3/2$ and $3$, respectively \citep{mat99}. Thus, the
energy per baryon increases by factors of $3.2$ and $2.8$ in these two
cases due to the concentration of energy in the fastest ejecta.

This phenomenon is even more dramatic in the context of relativistic
acceleration, where the postshock energy density is dominated by
internal energy rather than the rest mass of the fluid. For a strong
relativistic shock, the total postshock energy per unit rest mass,
evaluated in the fixed frame and neglecting terms of order
$\Gamma_s^{-2}$, is $4\Gamma_2^2/3=2\Gamma_s^2/3$ \citep{bla76}. If
each baryon maintained exactly this share of energy into the final
state of cold, free expansion, then $2\Gamma_s^2/3$ would also equal
its terminal Lorentz factor $\Gamma_f$. The theory of acceleration
behind an ultrarelativistic shock was treated approximately by
\cite{joh71}, who estimated $\Gamma_f \simeq
(\Gamma_s/\sqrt{2})^{1+\sqrt{3}}$ as a lower
limit.\footnote{\cite{joh71} showed $\Gamma_f$ would increase by a
multiplicative constant if the rest mass of the fluid were included in
the equation of state.} In this case, postshock acceleration increases
the energy per particle by a factor of approximately
$\Gamma_s^{0.73}$. As this is a large factor, the postshock
acceleration, and the concentration of energy in the fastest material,
are even more impressive in the relativistic case.

These considerations demonstrate that the acceleration of the fluid
after it is shocked is critical to the distribution of energy with
Lorentz factor in the ejecta. The character of planar postshock
acceleration is qualitatively different in the nonrelativistic case,
where the velocity increases by a fixed factor, from the relativistic
case, where the final Lorentz factor becomes a power of the postshock
value. However, in both cases there is a functional relationship
between $\Gamma_f\beta_f$ and $\Gamma_s\beta_s$. In reality there is a
dependence on the slope of the density distribution as well, but this
is quite weak and we are justified in ignoring it.  Hence our goal is
to find an approximate analytical relationship between the final
velocity of a fluid element and the velocity of the shock that struck
it: one that obeys the known limits and remains sufficiently accurate
in the trans-relativistic regime. We shall begin by examining postshock
acceleration in strictly planar geometry. Then we shall consider how
spherical expansion may reduce the terminal Lorentz factor of the
ejecta relative to the planar case.

\subsubsection{Planar Geometry} \label{PlanarPostshock}

\begin{figure}
\epsscale{1.0}
\plotone{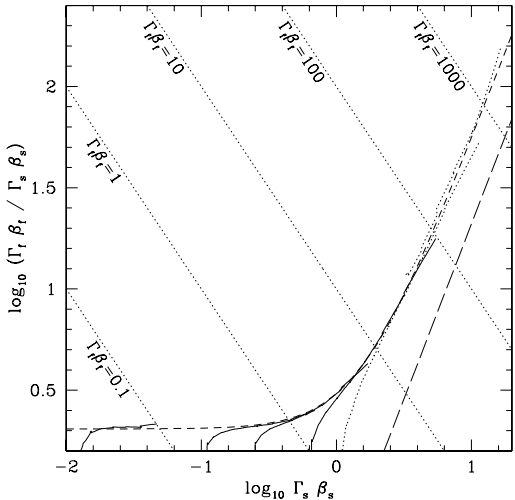}
\caption{Final fluid velocity ($\Gamma_f \beta_f$) in the {\it planar} 
free coasting state compared to the velocity of the shock ($\Gamma_s
\beta_s$) striking a particular fluid element, as a function of
$\Gamma_s \beta_s$. Numerical results for different initial energies
are shown by the heavy {\it solid} and {\it dotted} lines. The {\it
dotted} lines, at the highest energies, have the greatest systematic
uncertainties from corrections for incomplete convergence. The {\it
dashed} line is our expression for planar postshock acceleration
$\Gamma_f \beta_f / \Gamma_s \beta_s = C_{\rm nr} + (\Gamma_s\beta_s
)^{\sqrt{3}}$ (eq. [\ref{planar}]).  The inclined {\it long dashed}
line is the ultrarelativistic prediction of \citet{joh71},
$\Gamma_f=\Gamma_2^{1+\sqrt{3}}= 0.39 \Gamma_s^{1+\sqrt{3}}$, which we
find underestimates $\Gamma_f$ by a factor of $\sim2.6$.
\label{fig:planar}}
\end{figure}

We continue with the density distribution $\rho=\rho_h (1-r/R)^n\equiv
\rho_h x^n$, with $n=3$, which now describes a planar slab of material
of length $R$. We release shocks of varying strengths that propagate
down this density ramp. In regions significantly far away from the
launching region ($R-r\ll R$), and except for the effects of finite
resolution, the shock and postshock behavior ought to be the same
among all of these runs -- i.e., the function
$\Gamma_f\beta_f(\Gamma_s\beta_s)$ should have the same form. This
follows from the fact that a power law density distribution introduces
no scales of its own. By exploring the relation between shock and
postshock velocities in simulations of different initial shock
strength, we are able to discriminate this common behavior from the
artifacts of the particular initial conditions or of the finite
numerical resolution: see Figure \ref{fig:planar}.

For the purposes of numerical stability, we allow the fluid to expand
into a uniform medium of sufficiently low density that its effects on
the accelerating ejecta (a reverse shock) can be visualized and
excluded from consideration. Because the conversion of internal into
kinetic energy is quite slow in planar expansion, and because of
time-step constraints on our code, it was necessary to examine the
approach of mass elements toward their terminal velocities in order to
estimate how much acceleration would occur if they were allowed to
expand indefinitely. A correction to $\Gamma \beta$ based on the
extrapolation of internal energy to zero was made. Typical values of
this correction were $\lesssim 5-10\%$ for the lower energy runs ({\it
solid} lines in Figure \ref{fig:planar}). For the most relativistic
cases considered ({\it dotted} lines) the correction was $\sim 20 -
30\%$. This correction procedure was checked and validated against the
end state of a single simulation, allowed to expand out to a very
large distance, at which point the internal energy of the zones was
negligible. These corrected results were then further corrected for
incomplete numerical convergence, for both the shock and final
velocities, by considering the results as a function of the number of
simulation zones. The largest values of these corrections, again for
the most relativistic runs, were $\sim 50\%$.

We find that the following analytical formula captures the
relationship between shock and terminal velocities:
\begin{equation} \label{planar}
\frac{\Gamma_f \beta_f}{\Gamma_s \beta_s}\simeq C_{\rm nr} + (\Gamma_s \beta_s )^{\sqrt{3}}.
\end{equation}
For density distributions with $n\simeq 3$, $C_{\rm nr} = 2.03$
\citep{mat99}.  Comparing to numerical simulations, equation
(\ref{planar}) is accurate to $\pm10\%$ for $\Gamma_s
\beta_s\lesssim5$, corresponding to $\Gamma_f \beta_f\lesssim100$; and
$\pm25\%$ for $\Gamma_s
\beta_s\lesssim10$, corresponding to $\Gamma_f \beta_f\lesssim500$. 
The uncertainty is greatest for the most relativistic ejecta.  In the
ultrarelativistic limit, equation (\ref{planar}) tends towards the
scaling predicted by \citet{joh71}, but increased by a factor of
$\sqrt{2}^{1+\sqrt{3}}\simeq2.6$.

\subsubsection{Spherical geometry} \label{SphericalPostshock}

\begin{figure}
\epsscale{1.0}
\plotone{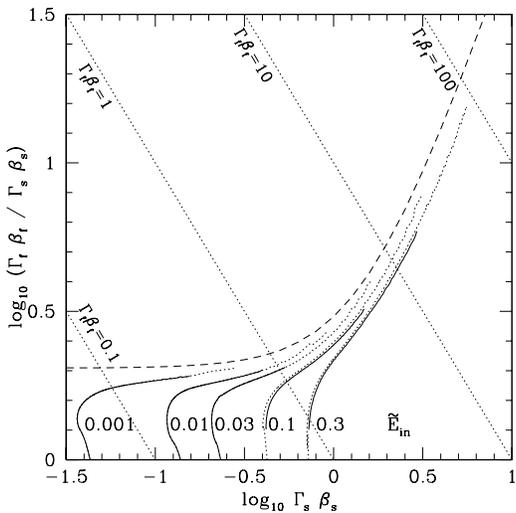}
\caption{Spherical postshock acceleration for simulations with
$\ein=0.001,0.01,0.03,0.1,0.3$. {\it Solid} lines are regions where we
have corrected for numerical convergence. {\it Dotted} lines are the
highest resolution results without such a convergence correction. The
{\it dashed} line shows the planar result from Figure \ref{fig:planar}.
\label{fig:spherical}}
\end{figure}

\begin{figure}
\epsscale{1.0}
\plotone{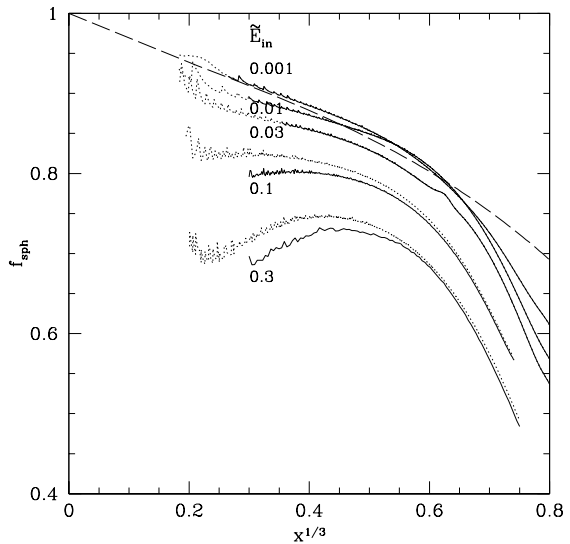}
\caption{Deviation of spherical postshock acceleration from the
planar limit (equation \ref{planar}) as a function of $x^{1/3}\equiv
(1-r/R)^{1/3}$, for simulations with a power-law initial density
distribution ($\rho\propto x^3$) and $\ein =
0.001,0.01,0.03,0.1,0.3$. The {\it solid} and {\it dotted} lines are
the same as in Figure \ref{fig:spherical}. The {\it long dashed} line
shows the nonrelativistic prediction for $f_{\rm sph}$ for an $n=3$
polytrope (eq. [\ref{spherical}]). For $x^{1/3}>0.6$ the chosen
initial conditions no longer approximate a polytrope, which explains
the discrepancy between theory and simulation in this region for
nonrelativistic ejecta. Uncertainties in the planar solution also
contribute to the accuracy of this prediction.
\label{fig:fspherical}}
\end{figure}

The above calculation assumes that planar symmetry holds throughout
the postshock acceleration phase of the ejecta. In fact, the planar,
nonrelativistic flow approaches its terminal velocity extremely
slowly: any fluid element must travel many times its initial depth
before acceleration is complete. Once the fluid has traveled a
distance of order the stellar radius, however, its volume is
significantly larger than in the planar approximation, and its pressure
is correspondingly less. This causes nonrelativistic acceleration to
cease at some radius, as originally suggested by
\citet{lit90}. Comparing to nonrelativistic numerical simulations,
\citet{mat99} find that the results of spherical expansion match the
results of planar acceleration, truncated at three stellar radii; this
is also consistent with the analytical theory of \citet{kaz92}. In
terms of the normalized depth $x\equiv 1-r/R$, \citet{mat99} find that
spherical truncation reduces the terminal velocity of a mass element
originally at depth $xR$ by the factor
\begin{equation}
f_{\rm sph, nr} (n=1.5) = 1 - 0.51x^{1/3} + 0.76x^{2/3} - 1.19x
\end{equation}
\begin{equation} 
f_{\rm sph, nr} (n=3) = 1 - 0.34x^{1/3} + 0.24x^{2/3} - 0.37x\label{spherical}
\end{equation}
for polytropes of these indices. These results on postshock
acceleration are valid for polytropic envelopes in an outer region
with $\mx\equiv 1-\mint \ll 1$, and $x^{1/3}\lesssim 0.8$. Deeper into
the envelope, polytropes differ in structure if they enclose different
core masses; this, along with changing dynamics deep within the
explosion, causes deviations from these common forms \citep{mat99}.

We have followed the postshock acceleration of ejecta from spherical
explosions of different energies, using for a progenitor the external
power law distribution $\rho=\rho_h (1-r/R)^3\equiv\rho_h x^3$, to examine how
spherical truncation, expressed via
\begin{equation}
f_{\rm sph} \equiv \frac{(\Gamma_f\beta_f)_{\rm spherical}}{(\Gamma_f\beta_f)_{\rm planar}},
\label{fsphdef}
\end{equation}
changes as velocities become more and more relativistic. Figures
\ref{fig:spherical} and \ref{fig:fspherical} compare the planar law
(eq. [\ref{planar}]) with spherical numerical results. The predicted
nonrelativistic truncation (eq. [\ref{spherical}]) is recovered for
$x^{1/3}<0.6$, where $\rho\propto x^3$ well approximates a
polytrope. At higher energies and velocities we observe more
significant truncation in the outermost ejecta. $\Gamma_f \beta_f$ is
reduced to $\sim 70\%$ of the predicted planar value in the most
relativistic case. We note this estimate is somewhat uncertain,
because of uncertainties in the planar solution to which we make
comparison.

For purposes of estimating the final energy distribution of the outer
($x<0.3, \:x^{1/3}<0.67$) ejecta we approximate $f_{\rm sph}$ as a
constant factor, independent of $x$, varying between $0.85$ and $0.7$
as $\ein$ is increased from 0.0 to 0.3. This behavior is
captured by
\begin{equation} 
f_{\rm sph}=0.85-\ein^{1/2}/4,
\label{fsph}
\end{equation}
which is approximately the value of $f_{\rm sph}$ at $x=0.125$. This
expression is accurate to $\pm 15\%$ at any given $x$, tending to
underestimate velocities at $x<0.125$, and overestimate those at
$x>0.125$. It has greater accuracy for predicting the mean value of
$f_{\rm sph}$ for $0.0<x<0.3$.

Combining expressions (\ref{relshockf}), (\ref{planar}), and
(\ref{fsph}), the final velocity of a mass element is
\begin{eqnarray}\label{FinalVelocityInTermsOfP}
\Gamma_f\beta_f=& (0.85-\ein^{1/2}/4) p(1 + p^2)^{0.12}\nonumber \\
~~~~~& \times[2.03 + p^{1.73}  (1 + p^2)^{0.21}]\nonumber\\
~~~~\simeq&(0.85-\ein^{1/2}/4) (1.80p^{5/6} + p^{2.82})^{6/5},~~
\end{eqnarray} 
where the second line approximates the first within $1.5\%$.  Either
approximation holds within $15\%$ only in the outermost $\sim 30\%$ of
the stellar radius, where equation (\ref{fsph}) is valid. 

In an outer layer ($x\lesssim 0.2$), the polytropic distribution
(eq. [\ref{eq:outerpolytrope}]) limits to the power-law form
(eq. [\ref{extpow}]). The mass and density are then related by
\citep{mat99}
\begin{equation} \label{defineFrho1}
\rho = f_\rho \frac{\Mej}{R^3} \mx^{1/\gammap}, ~~~(x\ll1), 
\end{equation}
where 
\begin{equation}
\gammap \equiv 1 + 1/n,
\end{equation}
\begin{equation} \label{defineFrho1_2}
f_\rho = \left(\frac{n+1}{4 \pi}\right)^{1/\gammap} \rhoht^{1/(n+1)}, 
\end{equation}
which is \citeauthor{mat99}'s $f_{\rho_0}(1)$, and we
have defined
\begin{equation} \label{definerhot}
\rhoht \equiv \frac{\rho_h R^3}{M_{\rm ej}}, 
\end{equation}
which is \citeauthor{mat99}'s parameter $\rho_1/\rho_\star$. 
For $n\simeq 3$, equation (\ref{defineFrho1}) is valid to within
$10\%$ for $r>0.96 R$ and to within $50\%$ for $r>0.75R$. The scaling
$\rho^\gammap \propto \mx$ arises from the hydrostatic relation
between mass and pressure in a thin subsurface layer.  

In this outer power-law region, $p$ can be evaluated by approximating
$r\simeq R$ and $\mint\simeq 1$ in equation (\ref{relshock}):
\begin{mathletters}\label{ExteriorPAppx}
\begin{eqnarray} 
p &\simeq& A {\ein}^{1/2} \left(\frac{\rho R^3}{\Mej}\right)^{-0.187}
\label{ExteriorPRho} \\ 
 &\simeq& A {\ein}^{1/2} \rhoht^{-0.187} x^{-0.187n} 
\label{ExteriorPX}\\ 
 &\simeq& A {\ein}^{1/2} f_\rho^{-0.187}\mx^{-0.187/\gammap}, \label{ExteriorPMx}
\end{eqnarray}
\end{mathletters}
where equations (\ref{ExteriorPX}) and (\ref{ExteriorPMx}) use (\ref{extpow})
and (\ref{defineFrho1}), respectively.  The error introduced by
adopting these power law approximations for $p$ is roughly $x$, and in
the outer fifth of the radius it is smaller than the error
introduced by $f_{\rm sph}$ in equation (\ref{fsph}).

As we have noted, equation (\ref{fsph}) for the parameter $f_{\rm
sph}$, and the power law representations for $p$ in equation
(\ref{ExteriorPAppx}), break down for $x\gtrsim 0.25$. As the shock
intensity $p$ increases outward near the surface of the star, this
implies that there exists a minimum value of $p$ (hence, from eq.
[\ref{FinalVelocityInTermsOfP}], a minimum final velocity) for which our
equations accurately predict the final velocity. To estimate this
criterion, we may set $x\lesssim 0.25$ and estimate $A \simeq 0.736$
in equation (\ref{ExteriorPX}) to find  
\begin{equation} \label{PValidityCriterion}
p \gtrsim  0.736 \exp(0.26n) {\ein}^{1/2} \rhoht^{-0.187}. 
\end{equation}

\subsection{Kinetic Energy Distribution of the Ejecta}\label{S:KEdist}

\begin{figure}
\epsscale{1.0}
\plotone{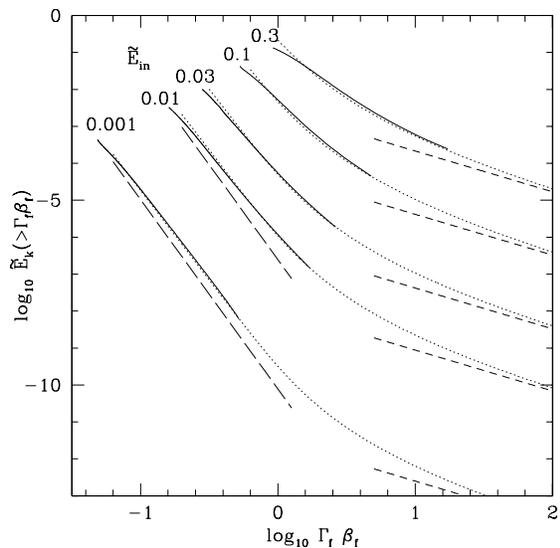}
\caption{Cumulative kinetic energy distribution of ejecta from a
progenitor with $\rho=\rho_h (1-r/R)^3$ for $\ein=0.001,0.01,0.03,
0.1,0.3$. {\it Solid} lines - converged results from numerical
simulation; {\it dotted} lines - analytic estimate
(eq. \ref{EkGeneral}); {\it dashed} lines - relativistic limit
(eq. \ref{EkLimitXR}); and {\it long-dashed} lines - nonrelativistic
limit (equation \ref{EkLimitNR}) shown for the two lowest energy
cases.
\label{fig:kean}}
\end{figure}

In order to estimate the observable properties of relativistic mass
ejection from astrophysical explosions, it will be useful to know the
kinetic energy in ejecta above some velocity
$\Ek(>\Gamma_f \beta_f)$. This information is available through the formulae
presented in \S\S \ref{S:shockprop} and \ref{shacc}. In this section,
we shall use the above results to present more convenient analytical
formulae for $\Ek(>\Gamma_f \beta_f)$. We restrict our attention to a shell of
material from the outermost layers of the star, whose mass is
negligible ($\mx\equiv 1-\mint\ll 1$) and through which the shock
accelerated. We approximate the spherical truncation with
eq. (\ref{fsph}), which tends to underestimate the kinetic energy in
relativistic ejecta, and overestimate the energy in nonrelativistic
ejecta, although never by more than 15\%.

The fraction of the rest mass energy transmitted into kinetic energy
at velocities above $\Gamma_f \beta_f$
is
\begin{equation} \label{finalke}
\ek(>\Gamma_f \beta_f) = \int_{0}^{\mx} (\Gamma_f - 1) d \mx',
\end{equation}
where $\mx$ refers to the ejecta traveling faster than
$\Gamma_f\beta_f$. 

Lacking a relativistic theory, \cite{mat99} estimated the energy in
relativistic ejecta as $M_{\rm rel} c^2$, where $M_{\rm rel}$ is the
mass in which an extrapolation of the nonrelativistic expressions
(ignoring spherical truncation) predicts $\beta_f>1$. Examining
equations (\ref{nonrelshock}), (\ref{relshock}) and (\ref{planar}), we find that this
definition corresponds to $p>1/C_{\rm nr}$:
\begin{equation}\label{defnMrel}
M_{\rm rel} \equiv \mx(p>1/C_{\rm nr}) \Mej. 
\end{equation}
Equations (\ref{relshockf}) and (\ref{planar}), together with
allowance for spherical truncation (eq. [\ref{fsphdef}]), 
show that the ejecta with mass $M_{\rm rel}$ expand with $\Gamma_f
\beta_f > 1.18f_{\rm sph}$; if $f_{\rm sph}\simeq 0.85$, as indicated
by equation (\ref{fsph}), this is $\Gamma_f\beta_f > 1.00$. $M_{\rm
rel}$ is thus an excellent estimate of the mass in relativistic
ejecta.

For a stellar progenitor with a
polytropic atmosphere, \cite{mat99} find
\begin{eqnarray} \label{mrel}
M_{\rm rel} &\simeq & {10^{28.5-4.8/n}} f_\rho^{-\gammap}
\left(\frac{A}{0.794}\right)^{5.35\gammap}
\nonumber\\ &\times & 
\left(\frac{\Ein}{10^{52}\:{\rm erg}}\right)^{2.67\gammap}
\left( \frac{M_{\rm ej}}{1\:{\rm M_{\odot}}}
\right)^{1-2.67\gammap}\:{\rm g},\nonumber\\
\end{eqnarray}
where $f_\rho$ is defined in equation (\ref{defineFrho1}) and we
have included the dependence on $A$; we shall present equivalent
expressions for progenitors with radiative outer envelopes below. 

Equation (\ref{mrel}) assumes that the shock velocity is a power law
in depth. This assumption is valid in the outer $\sim 25\%$ of the
stellar radius; equivalently, it holds when the value $p = 1/C_{\rm
nr}=0.49$, which corresponds to $M_{\rm rel}$, is above the lower
limit in equation (\ref{PValidityCriterion}). This requires
\begin{equation}\label{MrelValidityCriterion}
\ein \lesssim 0.44 \exp(-0.52n) \rhoht^{0.374}. 
\end{equation} 
If this condition is not satisfied, the value of $M_{\rm rel}$ in
equation (\ref{mrel}) will under predict the mass in material with
$p>0.49$, and equation (\ref{fsph}) will over predict its final
velocity. Even so, $M_{\rm rel}$ remains an accurate coefficient for
higher-velocity ejecta that do arise in the outer quarter of the
radius; c.f. equations (\ref{EkFromMrel2}) and (\ref{EkFromMrel3})
below.

To evaluate $M_{\rm rel} c^2$ as an estimate of the kinetic energy in
relativistic ejecta, and to determine the distribution of kinetic
energy with velocity, we must evaluate equation (\ref{finalke}). 
For nonrelativistic ejecta, equations (\ref{relshockf}), (\ref{relshock}),
(\ref{planar}), and (\ref{defineFrho1}) give $\beta_f\propto
\mx^{-\alphanr/\gammap}\propto\mx^{-0.187/\gammap}$, for $r\simeq R$ and $\mint\simeq1$. 
Since $\ek(>\Gamma_f \beta_f) = \int_0^{\mx} \beta_f^2
d\mxo/2$ for nonrelativistic ejecta, this implies that
\begin{equation}\label{EkIntermediateNR}
\ek(>\Gamma_f \beta_f) = \frac{\beta_f^{2} \mx }{2-0.75/\gammap}.
\end{equation}
For extremely relativistic ejecta, the same equations give $\Gamma_f
\propto \mx^{-0.634/\gammap}$, and since $\ek(>\Gamma_f
\beta_f) = \int_0^{ \mx} \Gamma_{f} d\mxo$ in this limit,
\begin{equation}\label{EkIntermediateR}
\ek(>\Gamma_f \beta_f) = \frac{\Gamma_f \mx  }{1-0.63/\gammap}. 
\end{equation}
Note that the energy above some final velocity $\Gamma_f
\beta_f$ is proportional to the mass $\mx$ in ejecta above
that velocity, quite generally. This follows from equations
(\ref{planar}) and (\ref{fsph}), which demonstrate that the final
velocity is essentially a universal function of the shock velocity for
a given element, modulo slight variations in $f_{\rm sph}$ not
captured in equation (\ref{fsph}). Therefore,
$\ek(>\Gamma_f \beta_f)$ is proportional to mass of ejecta that
was hit by a shock going faster than the corresponding shock velocity
$\Gamma_s \beta_s$. 

Eliminating $\mx$ from equations (\ref{EkIntermediateNR}) and
(\ref{EkIntermediateR}) by means of equations (\ref{relshockf}),
(\ref{planar}), (\ref{fsphdef}) and (\ref{defineFrho1}), we find in
the non- and ultrarelativistic limits
\begin{eqnarray}\label{EkLimitNR}
\ek(>\Gamma_f\beta_f) &\simeq &
\frac{(f_{\rm sph} C_{\rm nr} A \ein^{1/2})^{5.35\gammap}}
     {2 - 0.75/\gammap}
\nonumber \\& \times &
     f_\rho^{-\gammap} (\Gamma_f\beta_f)^{-(5.35\gammap-2)}\nonumber\\
&~& [\Gamma_f\beta_f \ll 1]
\end{eqnarray}
and 
\begin{eqnarray}\label{EkLimitXR}
\ek(>\Gamma_f\beta_f) &\simeq&
\frac{(f_{\rm sph}^{0.295} A \ein^{1/2})^{5.35\gammap}}
     {1 - 0.63/\gammap} 
\nonumber \\ &\times &
    f_\rho^{-\gammap} (\Gamma_f\beta_f)^{-(1.58\gammap-1)}
\nonumber \\
&~& [\Gamma_f\beta_f \gg 1]
\end{eqnarray}
respectively. One formula that attains each of these limits is 
\begin{equation}\label{EkGeneral}
\ek(>\Gamma_f\beta_f) = (A \ein^{1/2})^{5.35\gammap} f_\rho^{-\gamma_p} F(\Gamma_f \beta_f),
\end{equation}
where 
\begin{eqnarray} \label{F}
F(\Gamma_f\beta_f) &\equiv& \frac{(f_{\rm sph}^{1-0.705\beta_f}
C_{\rm nr}^{1-\beta_f})^{5.35\gammap}} {2-\beta_f
- (0.75-0.11\beta_f)/\gammap} \nonumber\\
&\times& \left[(\Gamma_f\beta_f)^{-(5.35-2/\gammap)/q}\right.\nonumber\\
&~&+\left.(\Gamma_f\beta_f)^{-(1.58 - 1/\gammap)/q} \right]^{q\gammap}.~~
\end{eqnarray}
We find that setting $q=4.1$ gives an excellent fit to the simulation
results, as Figure \ref{fig:kean} demonstrates. This fit is more
precise than deriving $E_k(>\Gamma_f\beta_f)$ from equations
(\ref{ExteriorPMx}), (\ref{relshockf}), (\ref{planar}), and
(\ref{fsph}), principally because details of the spherical truncation
are not captured in equation (\ref{fsph}). Note that for progenitor
density distributions that fall off rapidly near their edges, the
energy contained in ejecta above a particular velocity is quite
sensitive to the degree of shock and postshock acceleration. For
example a $10\%$ error in the final velocity caused by equation
(\ref{fsph}) can lead to a factor of two error in
$\ek(>\Gamma_f\beta_f)$ in the nonrelativistic limit. Variations in
the polytropic index $n$ may cause $\alpha_{\rm nr}$ to vary by a few
percent \citep{sak60}. Such differences may be amplified into large
variations in $\ek(>\Gamma_f\beta_f)$, typically $\sim 50\%$.
Note that the ultrarelativistic limit is
approached only very slowly, i.e. for $\Gamma_f \beta_f
\gtrsim100$. This is because the Lorentz factor of the shock,
$\Gamma_s$, needs to be relativistic for these limits to apply and
$\Gamma_f \beta_f \gg \Gamma_s \beta_s$ in the relativistic regime
(eq. \ref{planar}). Figure \ref{fig:kean} also illustrates that the
cumulative energy distribution deviates from equation
(\ref{EkGeneral}) in the lowest-velocity ejecta. This is due to the
failure of several of our assumptions in this material: $r$ is no
longer close to $R$, $f_{\rm sph}$ decreases at lower velocities, and
the dynamics changes once $\mx\ll 1$ is no longer valid; these issues
were treated in detail by \cite{mat99}.

Expression (\ref{EkGeneral}) is a remarkably simple estimate of the
fast and relativistic ejecta from a stellar explosion. In the
following sections, we shall use it to predict the outcomes of
explosions within radiative stellar envelopes, and of asymmetrical
explosions. We wish therefore to re-express (\ref{EkGeneral}) in a
couple of forms that will facilitate these investigations.

First, suppose one knows that the shock exceeds an intensity $p'$ in a
small fraction $\mx(>p')$ of the total ejecta. What will be the
kinetic energy in ejecta faster than $\Gamma_f\beta_f$, where
$\Gamma_f\beta_f$ is arbitrary?
Using equations (\ref{relshockf}),
(\ref{relshock}), and (\ref{defineFrho1}), we rewrite
(\ref{EkGeneral}) as
\begin{equation}\label{eq:EkMexp}
\ek(>\Gamma_f\beta_f)\simeq p'^{5.35\gammap} \mx(>p')
F(\Gamma_f\beta_f).
\end{equation} 
Here, the combination $p'^{5.35\gammap} \mx(>p')$ is a constant
independent of $p'$, so long as it is evaluated in the outermost $\sim
25\%$ of the radius (e.g., as in eq. \ref{ExteriorPMx}). This leads
equations (\ref{EkLimitNR}), (\ref{EkLimitXR}), and (\ref{EkGeneral})
to obey the scaling (in dimensional variables)
\begin{equation}\label{Ekscaling}
E_k(>\Gamma_f \beta_f) \propto [\Ein/(\Mej c^2)]^{2.67\gammap} \Mej c^2.
\end{equation}

Second, we may express $E_k(>\Gamma_f \beta_f)$ in terms of the rest
mass scale $M_{\rm rel}$ for relativistic ejecta.  Taking $f_{\rm sph}
= 0.85$ and $p'=1/C_{\rm nr}=0.49$, equation (\ref{eq:EkMexp}) gives
\begin{equation}\label{EkFromMrel2} 
E_k(>\Gamma_f\beta_f) \simeq \frac{F(\Gamma_f\beta_f)}{C_{\rm
nr}^{5.35\gammap}} M_{\rm rel} c^2. 
\end{equation} 
Evaluating this at the transition from nonrelativistic to relativistic
motion ($\Gamma_f\beta_f=1$), 
\begin{equation}\label{EkFromMrel3} 
E_k(>1) \simeq 1.23 \frac{n+1}{n+2.1} 
\exp\left(-\frac{1}{3.7n}\right) \left(\frac{f_{\rm
sph}}{0.85}\right)^{2.67\gammap} M_{\rm rel} c^2.  
\end{equation}
Hence  $E_k(>1) \simeq (0.71, 0.88, 0.94, 1.07)M_{\rm rel}c^2$ for
$n=(1.5, 3, 4, 9)$, respectively, justifying $M_{\rm rel} c^2$ as an
energy scale for relativistic motion. 

\subsection{Envelope Structure and the Efficiency of Fast Ejecta
Production} \label{S:whips} 
An accelerating shock front produces rapid ejecta in much the same
manner as a whip produces its crack: by concentrating a diminishing
fraction of its energy in an even more rapidly diminishing fraction of
the total mass, as a disturbance travels outward.  On Earth, whips are
fashioned to yield the sharpest report for the least effort, so we may
ask: What qualities in a stellar progenitor will maximize the energy
in ejecta above a given velocity?

To address this question, we rewrite equation (\ref{ExteriorPMx}) with
$p=0.49$, as
\begin{equation}\label{MrelForEfficiency}
\frac{M_{\rm rel}}{M_{\rm ej}} = \left[
\left(\frac{A}{0.736}\right)^2 \frac{2.23 \ein}{
f_\rho^{0.374}}\right]^{2.67\gammap}. 
\end{equation}
This ratio represents the efficiency with which a given explosion
creates relativistic ejecta. For a given value of $\ein$, what
properties of the ejected envelope enhance $M_{\rm rel}/M_{\rm ej}$?
Note that $A/0.736$ and $f_{\rho}$ are both roughly unity; moreover,
equation (\ref{ExteriorPMx}) is validated only for $\ein\lesssim
0.3$. For the mildly relativistic explosions to which
(\ref{MrelForEfficiency}) applies, it states that $M_{\rm rel}/M_{\rm
ej}$ is a small number raised to the power $2.67 \gammap$. The
relativistic yield of an explosion is therefore enhanced by making
$\gammap$ as small as possible, i.e., by making $n$ as large as
possible. For a fixed value of $\gammap$, the relativistic yield is
also enhanced by decreasing $f_\rho$, or equivalently, by decreasing
the outer density coefficient $\rhoht$. Both of these results can be
interpreted to say that more centrally concentrated envelopes are more
efficient at producing relativistic ejecta.

Second, note that the relativistic ejecta of a trans-relativistic
explosion are enhanced -- both in absolute terms and as a fraction of
the total -- if $\Ein$ is increased or $\Mej$ is reduced. For a given
progenitor mass, the relativistic yield is thus enhanced by allowing
as much of the star as possible to collapse. 

Equations (\ref{EkFromMrel2}) and (\ref{EkFromMrel3}) demonstrate that
the kinetic energy at {\em any} final velocity $\Gamma_f \beta_f$
(relativistic or otherwise) is proportional to $M_{\rm rel} c^2$, so
long as this ejecta originated from a thin outer layer in which the
shock accelerated; therefore, the above conclusions apply to fast
ejecta even when the maximum ejecta velocity is nonrelativistic.

Let us examine the criterion for relativistic ejecta to exist. The
alternative is that the shock diffuses through the surface before
$p$ reaches 0.49, the value for which $\Gamma_f\beta_f>1$. The optical
depth of a nonrelativistic shock front is $\tau_s \sim 1/p$
\citep{wea76}, whereas the optical depth of the layer with mass
$M_{\rm rel}$ is $\tau = \kappa M_{\rm 
rel}/(4\pi R^2)$, where $\kappa$ is the opacity. To contain the shock
front when $p=0.49$,  $\tau > \tau_s \simeq 2$, or 
\begin{equation} \label{RadiusConstraint}
R \lesssim 55 \left(\frac{\kappa}{0.34~{\rm cm^2g^{-1}}} 
\frac{M_{\rm rel} c^2}{10^{48}\rm erg}\right)^{1/2} R_\odot. 
\end{equation} 
This constraint on $R$, which is easily satisfied by the model CO6
considered below ($M_{\rm rel}c^2\sim 10^{48}$ erg; $R\sim
R_\odot/6$), becomes less restrictive as $M_{\rm rel}$ increases.

The mass $M_{\rm rel}$ in relativistic ejecta, and the corresponding
kinetic energy $E_k(>1)$, can be derived from equation (\ref{MrelForEfficiency})
using the equilibrium structures of stars with radiative outer
envelopes. In the case of a constant (e.g., electron scattering)
opacity, $n=3$ and
\begin{equation} \label{frhoThompson} 
f_\rho = 0.148  \left(\frac{M_\star}{\Mej}\right)^{1/4}
\frac{1-L_\star/L_{\rm edd}}{(L_\star/L_{\rm
edd})^{1/4}} 
\left(\frac{\mu}{0.62}\right)
\left(\frac{M_\star}{10~M_\odot}\right)^{1/2}
\end{equation} 
(\citealt{mat99}; note typo in their eq. 12), where $\mu$ is the mean
molecular weight in amu and $L_{\rm edd}$ is the Eddington limiting
luminosity. Using (\ref{MrelForEfficiency}) and (\ref{frhoThompson}),
we find that the fraction of the ejected mass that becomes
relativistic is 
\begin{eqnarray}\label{MrelMej-Chandra}
\frac{M_{\rm rel}}{M_{\rm ej}} &=& 430 \mu^{-4/3}\left(\frac{\Ein}{\Mej
c^2}\right)^{3.57} \left(\frac{A}{0.736}\right)^{7.13}
\nonumber \\ 
&~&\times \frac{M_{\rm ej}^{1/3} M_{\rm Ch}^{2/3}}{M_\star}
\frac{(L_\star/L_{\rm edd})^{1/3}} {(1-L_\star/L_{\rm edd})^{4/3}},
\end{eqnarray}
where $M_{\rm Ch}=1.434~M_\odot$ is the Chandrasekhar limiting mass
for an object with $\mu_e = 2$. As $L_\star$ increases relative to
$L_{\rm edd}$, the outer envelope becomes significantly more dilute
compared to the core; as we argued in the previous section, the
efficiency of relativistic ejection, $M_{\rm rel}/\Mej$, increases
correspondingly. This effect can be achieved, for instance, by
increasing the radiative opacity (hence lowering $L_{\rm edd}$):
intuitively, the envelope must expand to carry a constant flux despite
the higher opacity.

Equations (\ref{EkFromMrel3}) and (\ref{MrelMej-Chandra}) then give
\begin{eqnarray} \label{ErelThomson}
E_k(>1)&=& 0.88 M_{\rm rel}c^2 \nonumber \\
&\simeq&  (1.8, 2.6, 8.9)\times 10^{46} 
E_{52}^{3.57} L_5^{1/3} \nonumber \\ & \times & \left(\frac{M_\star}{6~\Msun}\right)^{-4/3} \left(1-\frac{L_\star}{L_{\rm edd}} \right)^{-4/3}~~~~
\nonumber \\ &\times&
\left(\frac{A}{0.736}\right)^{7.13}
\left(\frac{M_{\rm ej}}{4~\Msun}\right)^{-2.23}
{~\rm erg},~~~~~~
\end{eqnarray} 
using $f_{\rm sph} \simeq 0.85$, for stars with CO envelopes, He
envelopes, and H envelopes of solar composition, respectively. In this
equation, $ L_\star \equiv 10^5 L_5 L_\odot$ and $\Ein\equiv 10^{52}
E_{52}$ erg.

If the opacity in the outer envelope (the region containing $M_{\rm
rel}$) is dominated instead by bound-free or free-free transitions,
then $M_{\rm rel}$ will exceed the prediction of equation
(\ref{ErelThomson}). Provided radiation pressure is negligible, which
requires $L_\star \ll L_{\rm edd}$, the envelope will be a polytrope
with $n=3.25$ and 
\begin{equation}\label{frho-f.f.} 
f_\rho^{17} = \frac{2}{R} \left(\frac{\pi^2 a c}{51 \kappa_0
L_\star \Mej^2}\right)^2
\left(\frac{G M_\star \mu M_p}{2 \pi k_B}\right)^{15}. 
\end{equation}
Here $\kappa = \kappa_0 \rho T^{-3.5}$ is Kramers' opacity law; we
estimate $\kappa_0\simeq (37, 0.69, 1.24)\times 10^{24}~{\rm
cm^5~g^{-2}}$ in (CO, He, H) envelopes \citep{arp94,cla83}, by taking
both relevant Gaunt factors to be $\sim 0.85$ ($M_{\rm rel}$ depends only
weakly on this choice.)
The energy above $\Gamma_f\beta_f=1$ is, using $f_{\rm sph}=0.85$, 
\begin{eqnarray} \label{ErelKramers}
E_k(>1)&=& 0.90 M_{\rm rel} c^2 \nonumber \\
 &\simeq&
(3.2, 2.4, 6.3)\times 10^{46} 
E_{52}^{3.50} L_5^{2/13} \nonumber \\ &\times&
R_{10}^{1/13} \left(\frac{M_\star}{6~\Msun}\right)^{-15/13} \left(\frac{A}{0.736}\right)^{6.99}\nonumber\\
 & \times &  \left(\frac{M_{\rm
ej}}{4~\Msun}\right)^{-2.19}~{\rm erg},~~~~~
\end{eqnarray}
for (CO, He, H) envelopes, respectively, where $R \equiv 10^{10}
R_{10}$~cm.  In the Kramers layer, radiation pressure increases
outward as a fraction of the total; when it is no longer negligible,
$n$ increases toward seven \citep{cha39}; therefore, equation
(\ref{ErelKramers}) is an underestimate in stars with appreciable
$L_\star/L_{\rm edd}$. In general, the larger of 
(\ref{ErelThomson}) and (\ref{ErelKramers}) gives a lower limit for
the kinetic energy of relativistic material.

Let us examine the criteria for equations (\ref{ErelThomson}) and
(\ref{ErelKramers}) to correctly represent the energy in relativistic
ejecta. The first of these is given by condition
(\ref{MrelValidityCriterion}), which states that $M_{\rm rel}$ must
originate from the outer quarter of the radius, to correctly give
$E_k(>1)$. For (CO, He, H) electron-scattering atmospheres this
becomes, using eq. (\ref{frhoThompson}), 
\begin{eqnarray}\label{MrelCriterionThomson}
\ein &\lesssim& (0.12, 0.081, 0.021)\left(\frac{0.736}{A}\right)^{2}
\nonumber\\&\times& 
\left[\left(1-\frac{L_\star}{L_{\rm edd}}\right)^4
L_5^{-1} 
\left(\frac{M_\star}{6\: M_\odot}\right)^4  
\frac{4\:M_\odot}{\Mej}\right]^{0.374}. 
\end{eqnarray}
Using eq. (\ref{frho-f.f.}) to express the same condition for Kramers
atmospheres with negligible radiation pressure, 
\begin{eqnarray} \label{MrelCriterionKramers}
\ein &\lesssim& (0.087, 0.15, 0.037)\left(\frac{0.736}{A}\right)^{2}
\nonumber\\&\times& \left[L_5^{-1} R_{10}^{-1/2}
\left(\frac{M_\star}{6\: M_\odot}\right)^{15/2}   
\left(\frac{4\:M_\odot}{\Mej}\right)^2 \right]^{0.187}
\end{eqnarray}
in the same three cases. 
However, recall that $M_{\rm rel}$ can be used to predict the
distribution of ejecta at sufficiently relativistic velocities
(eqs. [\ref{EkFromMrel2}] and [\ref{EkFromMrel3}]), even when 
this condition indicates that it does not accurately determine
$E_k(>1)$. 

In contrast, there are no relativistic ejecta at all if the shock
breaks out of the star before $p>0.49$. This condition, expressed
above in equation (\ref{RadiusConstraint}), gives for Thomson
atmospheres 
\begin{eqnarray}\label{RCriterionThomson}
R_{10}&\lesssim& (43, 51, 122)
\left(\frac{A}{0.736}\right)^{3.57} L_5^{1/6} E_{52}^{1.78} 
\nonumber\\&\times&
\left(1-\frac{L_\star}{L_{\rm edd}}\right)^{-2/3} 
\left(\frac{6\: M_\odot}{M_\star}\right)^{2/3}   
\left(\frac{4\:M_\odot}{\Mej}\right)^{1.12}~~~
\end{eqnarray}
for (CO, He, H) envelopes. In the case that Kramers
opacity determines the structure of the envelope, it is still 
electron-scattering opacity that controls the width of the
shock. Therefore, equations (\ref{RadiusConstraint}) and
(\ref{frho-f.f.}) give 
\begin{eqnarray} \label{RCriterionKramers}
R_{10}  &\lesssim& (55, 48, 101) \left(\frac{A}{0.736}\right)^{3.64}
\nonumber\\&\times&
L_5^{2/25} E_{52}^{1.82} 
\left(\frac{6\: M_\odot}{M_\star}\right)^{3/5}   
\left(\frac{4\:M_\odot}{\Mej}\right)^{1.14}. 
\end{eqnarray} 

\subsection{Aspherical Explosions}\label{S:aspherical}
In equation (\ref{Ekscaling}), we have shown that the energy above
some velocity scales as $\Ein^{2.67\gammap}
\Mej^{1-2.67\gammap}$. Although our results so far have been limited
to spherical or planar symmetry, this scaling suggests that
asymmetries in the initial explosion may be exaggerated in the angular
distribution of relativistic ejecta. The dynamics of non-spherical
explosions are more appropriately addressed using multidimensional
numerical simulations
\citep[e.g.,][]{che89,mac99,hof99,alo00}. Because such simulations
cannot afford the fine surface zoning necessary to capture the
dynamics of shock acceleration, we wish to answer the following
questions: Under what conditions may the theory developed here for
spherical explosions be applied to aspherical explosions? What is the
appropriate correspondence between our formulae and the outcome of an
aspherical explosion?

To address the first question, it suffices to note that, during the
terminal acceleration of a nonrelativistic shock and the subsequent
postshock expansion \citep{sak60,mat99}, perturbations in the
postshock flow can only propagate laterally a limited distance: if the
shock has reached a depth $x$, then an acoustic wave launched from
some point on the shock surface can only affect a region within a few
times $x$ to either side. We arrive at this conclusion by
examining the self-similar solution presented by \cite{mat99} for the
planar version of this problem. Equation (A2) of their appendix allows
us to calculate the trajectory of a mass shell from the time it is hit
by the shock until the final state of free expansion. Given this
solution, their equation (A1) gives the history of the shell's
density, and hence, for adiabatic flow, its sound speed $c_s(t)$. The
integral of $c_s(t) dt$ gives the distance that a sound wave will
travel along this mass shell during the entire period of postshock
acceleration; we find $\int_{t_0(x)}^\infty c_s(t) dt = 1.17x$
and $2.00 x$ for $n=3$ and $n=1.5$, respectively, where $t_0(x)$ is
the time at which the shock was at the shell's initial depth $x$.

This indicates that the flow will remain planar as long as variations
in the position and strength of the shock front occur on length scales
large compared to the current depth, so that significantly different
patches of the surface are out of acoustical contact.  This becomes
even more true if the shock is relativistic, for then the lateral
region of influence is reduced in extent by a factor of about
$\Gamma_s$ by time dilation
\citep[e.g.,][]{1979ApJ...233..831S}. Likewise, the onset of spherical
expansion causes $c_s(t)$ to drop faster than the planar solution
would predict, further restricting the lateral acoustic range.  In
each direction this flatness criterion must be satisfied at some point
as the shock approaches the stellar surface; after that, the motion is
essentially planar. In reality, there are likely to be variations in
the effective value of the spherical expansion factor $f_{\rm sph}$,
if lateral expansion sets in on the correlation scale of the flow
rather than on the radial scale length $R$; this would have the effect
of changing the appropriate value of $x^{1/3}$. However, equation
(\ref{fsph}) shows that this variation can be ignored within the
accuracy of our theory.

Secondly, what is the angular distribution of ejecta from an
aspherical explosion? Sufficiently close to the outside of the star
that the above criterion is satisfied, each region of the surface
evolves independently of other regions separated by more than a few
times the initial depth. 
Only one parameter controls the evolution of each region: the initial
shock velocity.

Suppose that a numerical simulation has produced a shock whose
intensity is $p'$ (according to equation [\ref{relshock}]) below a
region with $d\mx(>p')/d A$. Equation (\ref{eq:EkMexp}) can be applied
to this patch of the surface, provided we replace the spherical
variables $\ek(>\Gamma_f\beta_f)$ and $\mx(r_{p'})$ with their
differentials:
\begin{equation}\label{eq:asphYp} 
\frac{d \ek(>\Gamma_f\beta_f)}{d A} \simeq p'^{5.35\gammap} F(\Gamma_f \beta_f) \frac{d\mx(>p')}{dA}; 
\end{equation} 
this emission is in the normal direction. Note that here, as in
(\ref{eq:EkMexp}), $p'$ is a reference value of $p$ and is not related
to $\Gamma_f \beta_f$ by equation (\ref{FinalVelocityInTermsOfP}).  To
gain insight into the implications of equation (\ref{eq:asphYp}) for a
given explosion, we make recourse to the so-called {\em sector
approximation}. Here, non-radial motions are ignored and each angular
sector is assumed to expand as if it were part of a spherical point
explosion with the local distributions of mass ($d\Mej/d\Omega$)
and energy ($d\Ein/d\Omega$).  Equation (\ref{eq:asphYp}), which can
be re-expressed in terms of $d\Omega$ instead of $dA$, then states
that the energy per unit solid angle above some velocity scales as
$(d\Ein/d\Omega)^{2.67\gammap} (d\Mej/d\Omega)^{1-2.67\gammap}$. In
the simplest case, the energy and mass are those in a sector of a
spherical explosion; in this case, the energy per unit area (or per
unit solid angle) above some reference velocity is unchanged. In the
opposite case in which all the energy is confined within some small
opening angle $\Omega_0$, the ejecta energy above some reference
velocity is increased by a factor
$(4\pi/\Omega_0)^{2.67\gammap}$. This result gives a qualitative
indication that aspherical explosions may appear far more intense (for
the fortunate viewer) than spherical ones. Moreover, this effect is
enhanced in an aspherical explosion in which the energy injection is
concentrated in directions in which $d\Mej /d\Omega$ is lower than
average.

Lastly, it should be noted that \cite{1990ApJ...359..463C} and
\cite{1994ApJ...435..815L} have identified an instability of 
nonrelativistic, accelerating shock fronts, in which the amplitude of
the perturbation doubles every four density scale heights; it 
saturates in a state where density is perturbed by a factor $2.5$ and
pressure by a factor $1.3$, and oblique weak shocks follow the leading
shock. These results were obtained for an exponential density
profile. However, a shock in a power-law density stratification can be
expected to behave similarly, except that the relevant perturbation
wavelength (about two density scale heights) will shrink in proportion
to the distance to the surface, leaving behind a spectrum of artifacts
in the postshock flow.  To estimate whether the instability will
saturate in a given star, note that the amplitude varies as
$\rho^{-1/6.9}$ according to \cite{1994ApJ...435..815L}, and therefore
as $\beta_s^{0.78}$. Because the shock's velocity is typically
$\beta_s\sim \ein^{1/2}$ when it begins to accelerate \citep{mat99},
the density perturbation will be enhanced by a factor $\sim
\ein^{-0.39}$ before the shock becomes relativistic (i.e., before
$p=1$). For the model of SN 1998bw described below, for which
$\ein\simeq 0.003$, this corresponds to an amplification of one
order of magnitude.

\subsection{Small Ejected Masses and the Role of
Gravity}\label{S:WD/NS-Theory} 

Thus far, we have considered only explosions caused by the collapse of
a stellar core to a much denser state; this has allowed us to neglect
the binding energy of the ejected envelope, and to use equations
(e.g., eq. [\ref{relshock}]) that assume a strong point explosion. In
this section we address the opposite limit: mass ejection by a shock
whose total energy is small compared to the binding energy of the
stellar envelope. This will allow us to analyze analytically the
ejecta from collapsing compact objects. 

In this limit, the shock is weak as it crosses most of the envelope,
because the thermal energy of a hydrostatic envelope is comparable to
its binding energy. Near the surface of the star, however, the
temperature is proportional to the depth. A shock is strong if it is
capable of ejecting material ($\beta_f c>v_{\rm esc}$, the escape
velocity), as the sound speed is $\sim x^{1/2} v_{\rm esc}$. Weaker
shocks are slower at any reference depth; but because they become
comparable to $v_{\rm esc}$ closer to the surface, they are stronger
(compared to the sound speed) at the point of ejection -- so long as
this occurs where $x\ll 1$.  Although we cannot use equation
(\ref{relshock}) to predict the shock velocity from the parameters of
the explosion, we can nevertheless apply our formulae for shock and
postshock acceleration, and use $v_{\rm esc}$ to set the energy scale
of what emerges.

We assume that the material of interest undergoes all of its
acceleration in the vicinity of the stellar surface; then, the ejecta
undergo a brief phase of acceleration followed by ballistic motion, in
which they either escape ($\beta_f c > v_{\rm esc}$) or fall back. As
relativistic stars are unstable, $E_{\rm grav}$(env)$<M_{\rm env}c^2$,
so the explosions under consideration are necessarily
nonrelativistic; likewise, any relativistic ejecta must escape, as
$v_{\rm esc}$ can only be a fraction of $c$.

We take $p$ to be given at some point where the shock has accelerated
and become strong; thereafter, the shock (eq. [\ref{relshockf}]) and
postshock (eqs. [\ref{planar}] and [\ref{fsphdef}]) dynamics follow
the theory we have developed.  Under the assumption of ballistic
motion, we may simply subtract the specific binding energy,
$GM_\star/(c^2R) = v_{\rm esc}^2/(2c)$, from the predicted values of
$\Gamma_f$; equivalently, the total kinetic energy in an external mass
$\Mx$ is reduced by $\Mx v_{\rm esc}^2/2$.  As this offset is less
than the rest energy $\Mx c^2$, it clearly has no effect on ejecta
with $\Gamma_f\gg 1$.  However, by introducing a cutoff velocity (the
escape velocity $v_{\rm esc}$), it allows us to find the total kinetic
energy for all the ejecta that escape. Setting $\Mx = \Mej$ and
$\beta_f c=v_{\rm esc}$ in equation (\ref{EkIntermediateNR}) and
subtracting $v_{\rm esc}^2 \Mej/2$ from the result, we find
\begin{equation} \label{gravesc}
E_k \simeq \frac{3n}{5n+8}
\frac{GM_\star}{R_\star} \Mej.
\end{equation} 
Insofar as Newtonian gravity remains valid, this expression is
accurate to within $3\%$ even for neutron stars. 

How is the net kinetic energy in equation (\ref{gravesc}) distributed
in velocity?  For this, let us identify a shock intensity $p_{\rm
esc}$ that gives $\beta_f c= v_{\rm esc}$ via equations
(\ref{relshockf}), (\ref{planar}), (\ref{fsphdef}), and (\ref{fsph}). If we set
$p'=p_{\rm esc}$ in equation (\ref{eq:EkMexp}), then $\mx(>p_{\rm
esc})=1$, as $p_{\rm esc}$ determines $\Mej$; thus,
\begin{equation}\label{Ek-WD/NS}
E_k(>\Gamma_f \beta_f) \simeq  p_{\rm esc}^{5.35\gammap} F(\Gamma_f\beta_f)
\Mej c^2, 
\end{equation} 
which is valid so long as the binding energy may be neglected, i.e.,
$\Gamma_f\beta_f \gg v_{\rm esc}/c$. 

For a nonrelativistic escape velocity, $v_{\rm esc} = C_{\rm nr}
f_{\rm sph} p_{\rm esc}$.  In equation (\ref{Ek-WD/NS}), when
$\beta_f\ll 1$, the kinetic energy above some $\Gamma_f\beta_f$ is
independent of $C_{\rm nr}$ and $f_{\rm sph}$, as their influences on
$p_{\rm esc}^{5.35\gammap}$ and $F(\Gamma_f\beta_f)$ cancel.  This is
to be expected, as $v_{\rm esc}$ sets the velocity scale and $\Mej
v_{\rm esc}^2$ sets the energy scale (eq. [\ref{gravesc}]), and
shock acceleration introduces no scales on its own, so $E_k/(\Mej
v_{\rm esc}^2)$ should be a pure function of $\beta_f c/v_{\rm
esc}$. These arguments no longer hold for relativistic ejecta, because
the ratio $v_{\rm esc}/c$ becomes important, and indeed, equation
(\ref{Ek-WD/NS}) depends explicitly on $C_{\rm nr}$ and $f_{\rm sph}$
when $\Gamma_f\beta_f\gtrsim 1$. With $C_{\rm nr} = 2.03$ and $f_{\rm
sph}\simeq 0.85$, (\ref{Ek-WD/NS}) becomes 
\begin{mathletters}
\label{Ek-WD/NS-evaluated}
\begin{eqnarray} 
\label{Ek-WD-a}
E_k(>\Gamma_f \beta_f) &\simeq& 10^{43.53 - 9.72/n} \left(\frac{0.85}{f_{\rm sph}}\right)^{5.35\gammap}
\nonumber \\ &\times& 
\left(\frac{M_\star}{1.4~\Msun} \frac{6000~{\rm
km}}{R}\right)^{2.67\gammap}
\nonumber \\ &\times& 
\left(\frac{\Mej}{0.1~\Msun}\right) F(\Gamma_f\beta_f)~{\rm erg}~~~~~~~
\\&\simeq& 
10^{50.96 - 2.29/n}
\left(\frac{0.85}{f_{\rm sph}}\right)^{5.35\gammap} 
\nonumber \\ &\times& 
\left(\frac{M_\star}{1.4~\Msun} \frac{10~{\rm
km}}{R}\right)^{2.67\gammap}
\nonumber \\ &\times& 
\left(\frac{\Mej}{0.1~\Msun}\right) F(\Gamma_f\beta_f)~{\rm erg}.~~~~~~
\label{Ek-NS-b}
\end{eqnarray} \end{mathletters} 
White dwarfs are sufficiently nonrelativistic that equation
(\ref{Ek-WD-a}) is a good estimate for the energetics of their
ejecta. For neutron stars, the escape velocity is high enough that
equation (\ref{Ek-NS-b}) overestimates the kinetic energy of their
ejecta by $15-30\%$. But, the $\sim 15\%$ uncertainty in $f_{\rm sph}$
causes an additional uncertainty in $E_k(>\Gamma_f \beta_f)$ of about
a factor of two, for the relativistic ejecta alone. 

\subsubsection{Characteristic Velocities of
Explosions}\label{S:CharacteristicVelocities} 

Equation (\ref{gravesc}) implies that, when ejecta are flung
ballistically from the surface of an object and $v_{\rm esc}$
determines any fall-back, their mean velocity closely resembles $v_{\rm
esc}$, and hence $ E_k \sim \Mej v_{\rm esc}^2/2$. Alternatively, if
$\Ein$ is known to exceed $\Mej v_{\rm esc}^2/2$, for
instance if $\Ein$ exceeds the binding energy of the entire stellar
envelope, then it is not consistent to assume that only a small outer
portion will be ejected.

In general, the explosion energy is determined at the base of the
ejecta, from which it is transferred to the rest in a
blast wave. Typically, the velocity scale at the base of the ejecta is
the escape velocity of the forming remnant, or the infall velocity of
material accreting onto it, as is clearly true for an accretion shock
that ejects material by accelerating outward.
\cite{1987ApJ...318L..57B} and \cite{woo92} identify a neutrino-driven
wind, with a velocity $\sim 0.35 c$ (close to the escape velocity from
a newborn neutron star), as an important possible component of
core-collapse supernovae and of accretion-induced collapse in white
dwarfs; this would set an upper limit for the characteristic velocity
at the base of the ejecta. In a thermonuclear explosion, the energy
per nucleon prescribes a (nonrelativistic) characteristic velocity.
If the remnant is a black hole, the characteristic velocity may
resemble that of the last stable orbit (mildly relativistic). However,
the existence of relativistic radio jets and pulsar winds
indicates that magnetically-dominated outflows violate this
estimate (and are typically asymmetrical). But, the postshock pressure
of a blast wave driven by such a wind cannot exceed the wind's ram
pressure; this gives a lower characteristic velocity than that of the
wind itself. 

One implication of this inner velocity scale is that equation
(\ref{relshockf}) becomes invalid near the base of the ejecta, where
it predicts higher velocities because it assumes a point explosion.
Another implication is that the average ejecta velocity defined by
$\bar\Gamma=1+\ein$ cannot exceed this characteristic inner
velocity. However, the mean ejecta velocity may be much lower, because
of neutrino losses, photodissociation, and work done against gravity,
or simply because a relatively massive envelope is swept up.

\section{Astrophysical Applications} \label{S:applications} 

\begin{deluxetable}{ccccc} 
\small \tablecaption{Parameters of Density Models\label{tab:eqpara}}
\tablewidth{0pt} 
\tablehead{ \colhead{Model} & \colhead{$\ein$} &
\colhead{$A$} & \colhead{$n$} & \colhead{$f_\rho$}\\ }
\startdata 
$\rho=\rho_h x^3$ & $\leq 0.03$ & 0.68 & 3 & 0.63\\
$\rho=\rho_h x^3$ & $0.1$ & 0.72\tablenotemark{\dagger} & 3 & 0.63\\ 
$\rho=\rho_h x^3$ & $0.3$ & 0.74\tablenotemark{\dagger} & 3 & 0.63\\ 
98bw ($\mx\gtrsim2\times 10^{-5}$) & 0.003 & 0.736 & 9 & 1.4\\ 
98bw ($\mx\lesssim2\times 10^{-5}$) & 0.003 & 0.736 & 4 & 0.3\\ 
WD & 0.01-0.1 & 0.705 & 1.9 & 0.3\\ 
NS ($\mx\gtrsim10^{-4}$)& 0.01-0.1 & 0.68 & 1.9 & 0.8\\
NS ($\mx\lesssim10^{-4}$)& 0.01-0.1 & 0.68 & 4 & 0.9 \\ 
\enddata
\tablenotetext{\dagger}{These values, determined by numerical simulation, 
are $\sim 5-10\%$ higher than predicted by eqs. (\ref{Anr}), (\ref{sigma}) 
and (\ref{krhoJl}).} 
\end{deluxetable}

\begin{figure}
\epsscale{1.0}
\plotone{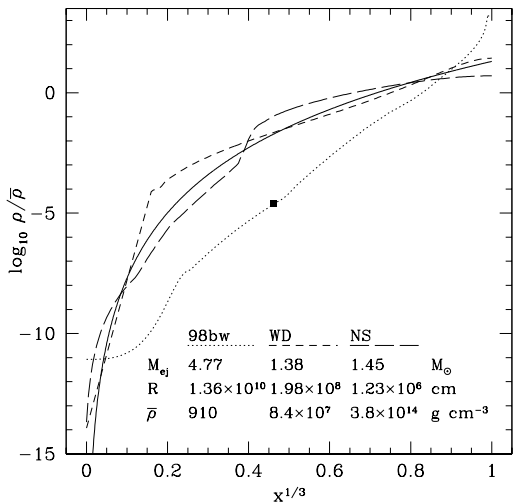}
\caption{Comparison of initial density distributions of our models of 
SN 1998bw ({\it dotted}) , white dwarfs (WD, {\it dashed}), and
neutron stars (NS, {\it long-dashed}). The {\it solid} line is the
idealized $\rho=\rho_h x^3$ distribution from \S\ref{S:theory}. The
{\it square} marks the location of the outer most zone considered by
\citet{woo99}. Beyond this we extrapolate with polytropic models out to
the circumstellar wind. A remnant mass of $1.78\:{\rm M_{\odot}}$ has
been removed from the center of this distribution, leaving $4.77\:{\rm
M_{\odot}}$.
\label{fig:denall}}
\end{figure}

We now consider four astrophysical examples where relativistic ejecta
may result from shock and postshock acceleration caused by an violent
explosion in a centrally concentrated density distribution. We use
numerical simulations in tandem with our analytic approximations
developed in \S\ref{S:theory}. First we present a detailed model of an
energetic carbon-oxygen core-collapse supernova, designed to model SN
1998bw, which has been associated with GRB 980425 \citep{gal98}. The
progenitor density distribution is based on a model kindly provided by
Stan Woosley. Next we make a simple estimate of the ejecta properties
of much more extreme ``hypernova'' explosions in such a progenitor,
and explore their ability to account for cosmological GRBs.  Finally
we present models of the expulsion of the outer layers of white dwarfs
(WDs) and neutron stars (NSs), that undergo a hypothesized core
collapse to a more compact state. The WD progenitor was generously
supplied by Lee Lindblom. The density distributions of all these
models are shown in Figure \ref{fig:denall}.

\subsection{SN 1998bw and GRB 980425} \label{S:98bw}

SN 1998bw was a very unusual supernova. Occurring in the nearby spiral
galaxy ESO 184-G82 at $z=0.0085$, equivalent to $38 \:{\rm Mpc}$ for
$H_0=65\:{\rm km\:s^{-1}\:Mpc^{-1}}$ \citep{tin98}, it is one of the
most luminous and earliest peaking Type Ib/c radio supernovae ever
observed \citep{li99,wei00}: its maximum radio luminosity was $\sim
4\times 10^{38}\:{\rm ergs\:s^{-1}}$, occurring $12\pm2$ days after
core-collapse \citep{kul98}. Various authors have derived differing
shock expansion velocities from synchrotron models of this emission,
finding $\beta_s\sim 0.3-0.8$ about 12 days after the start of the
supernova \citep{kul98,wax99,li99}.
The more relativistic the expansion, the greater the chance that
$\gamma$-rays could have been produced immediately after the star
exploded. We shall consider these models further in
\S\ref{S:radio}, where we compare them to our predictions from the
distribution of high-velocity ejecta.

Because of the lack of H and \ion{He}{1} features in its optical spectrum, SN
1998bw has been classified as a peculiar Type Ic event
\citep{pat98}. The optical spectra also reveal very broad emission
line blends of \ion{Fe}{2}, \ion{Si}{2}, \ion{O}{1}, and \ion{Ca}{2},
implying very high expansion velocities $\sim0.2c$ about a week after
the explosion. The peak optical luminosity ($\sim 1 \times
10^{43}\:{\rm ergs\:s^{-1}}$) is about ten times brighter than typical
SNe Ib/Ic. Modeling SN 1998bw's spectra and light curve under the
assumption of a spherically symmetric explosion of a C+O star, with
$\Mej \sim (5 - 10)\:{\rm M_{\odot}}$, requires large synthesized Ni
masses ($\gtrsim 0.5\:{\rm M_{\odot}}$) and explosion kinetic energies
($\sim (2 - 5) \times 10^{52}\:{\rm ergs}$) \citep{iwa98, woo99,
sol00, nak00}, again about an order of magnitude greater than from
typical supernovae. However, \citet{hof99} have presented asymmetric
explosion models that require smaller energies.

GRB 980425 also has some unusual properties distinguishing it from
other bursts. Observed by the BATSE detector on the {\it Compton Gamma
Ray Observatory} \citep{blo98} and the {\it Beppo-SAX} satellite
\citep{sof98}, no emission was seen above $300\:{\rm keV}$, making this
an example of the ``no high energy'' bursts that compose $\sim25\%$ of
the BATSE sample. The burst profile was smooth and single peaked,
showing little internal variability. At most, only seven other bursts
($0.5\%$ of the total GRB population) display such profiles
\citep{1999ApJ...518..901N}. The burst duration, $t_b$, was about
35~s, with the count rate of photons above 40~keV peaking four or five
seconds before the count rate of photons below 26~keV
\citep{2000ApJ...536..778P}. \citeauthor{2000ApJ...536..778P} also report 
that the supernova is coincident with an X-ray source that fades by a
factor of two over the first six months.

However, the most intriguing aspect of the unusual events SN 1998bw
and GRB 980425 is their spatial and temporal association
\citep{gal98}, occurring with $\lesssim 10^{-4}$ chance probability
according to {\it a posteriori} statistics. Their peculiarities only
serve to strengthen the suspicion that they are related. Given
association, the $\gamma$-ray energy of the burst was
$E_{\gamma}=(8.1\pm1.0) \times10^{47}\:{\rm ergs}$, equivalent to $\sim
10^{27}\:{\rm g}\:c^2$. Such a weak burst, thousands to millions of
times fainter than the inferred isotropic energies of cosmological
bursts, implies a separate GRB population. Corroborating this
suspicion, \cite{2000ApJ...534..248N} have shown that there is a power
law anti-correlation between intrinsic luminosity and hard-to-soft
time lags among six other bursts with known redshifts, but that
GRB 980425, if associated with SN 1998bw, is several hundred times
weaker than this law would predict. Moreover,
\cite{1999ApJ...520...54H} have shown that if the rate at which GRBs
occur within a galaxy is proportional to its luminosity, then the odds
of detecting a burst at redshift $z=0.0085$ in the current sample of
GRBs with known redshifts are less than one in $10^4$. Non-Euclidean
number count statistics limit the fraction of observed bursts
resulting from this mechanism to $\lesssim 10\%$ \citep{blo98}.

\subsubsection{Relativistic Ejecta from SN 1998bw}\label{SS:98bw}

\begin{deluxetable}{cc} 
\small
\tablecaption{SN 1998bw Model Parameters\label{tab:98bw}}
\tablewidth{0pt}
\tablehead{
}
\startdata
$M_*$ & 6.55 ${\rm M_{\odot}}$\\
$M_{\rm rem}$ & 1.78 ${\rm M_{\odot}}$\\
$M_{\rm ej}$ & 4.77 ${\rm M_{\odot}}$\\
$\Ein$ & $2.8 \times 10^{52}\:{\rm ergs}$\\
$R$ & $1.40 \times 10^{10}\:{\rm cm}$\\
\enddata
\end{deluxetable}

\begin{figure}
\epsscale{1.0}
\plotone{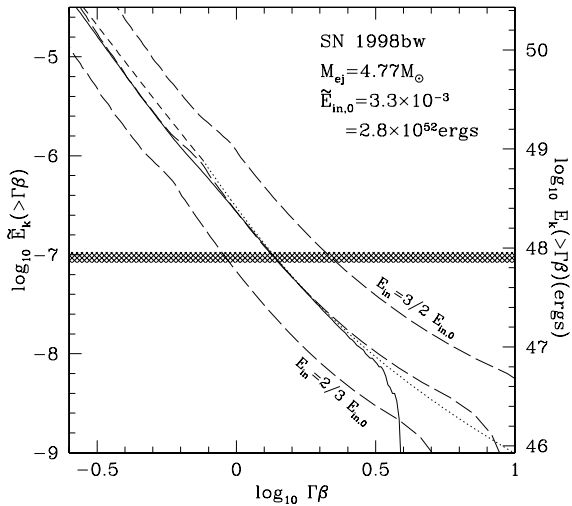}
\caption{Cumulative kinetic energy distribution of ejecta for SN
1998bw from simulation ({\it solid} line) and theory: the
central {\it long dashed} line is numerical integration of equation
(\ref{finalke}) over the progenitor density distribution, with
$\Gamma_f$ predicted from equations (\ref{relshock}),
(\ref{relshockf}), (\ref{planar}) and (\ref{fsph}); the upper and
lower {\it long dashed} lines show the effect of increasing and
decreasing the initial energy by a factor of 1.5, for this method;
predictions from equation (\ref{EkGeneral}) are shown by the {\it
dotted} ($n=4$, $f_\rho=0.3$, fit to outer ejecta) and {\it
dashed} ($n=9$, $f_\rho=1.4$, fit to inner ejecta)
lines. The {\it hatched} area shows the uncertainties in the
observed $\gamma$-ray energy of GRB 980425.
\label{fig:ke98bw}}
\end{figure}

\begin{figure}
\epsscale{1.0}
\plotone{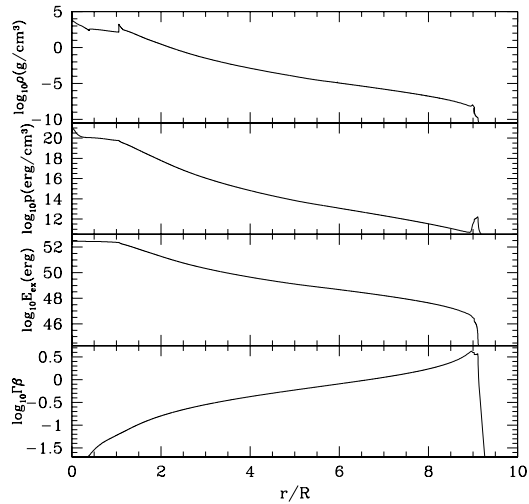}
\caption{Density, pressure, energy ($E_{\rm ex}$), and specific momentum
($\Gamma\beta$) of the ejecta from simulated supernova explosion CO6,
after expansion to 9 $R$ into a wind whose mass loss rate is
$\dot{M}_w=10^{-5}\:{\rm M_{\odot}\:yr^{-1}}$. $E_{\rm ex}$ is the
kinetic + thermal energy external to a particular location. The
reverse shock due to the circumstellar interaction is visible in the
distributions of density, pressure and velocity.
\label{fig:displayejecta}}
\end{figure}

We wish to examine in detail the question of whether the relativistic
ejecta from SN 1998bw were plausibly sufficient to power GRB 980425.
Previous estimates of the mass and energy associated with relativistic
supernova ejecta have extrapolated nonrelativistic results to the
relativistic regime; 
in this manner, \cite{mat99} predicted sufficient energy in
relativistic motion to account for the gamma-ray burst.
\citet{woo99} simulated shock break-out from a
variety of progenitor models for SN 1998bw with ejected masses in the
range $5-12\:{\rm M_{\odot}}$ and explosion energies $(4-28)\times
10^{51}\:{\rm ergs}$.  Our investigation of SN~1998bw 
will be restricted to \citeauthor{woo99}'s model CO6,
which Stan Woosley has kindly provided us. This is the bare
carbon-oxygen core ($\sim6.6~\Msun$) of a star that was initially more
massive ($\sim25~\Msun$); its pre-explosion luminosity is $6.6 \times
10^{38}\:{\rm ergs\:s^{-1}}$. The supernova forms a neutron star of
$1.78\:{\rm M_\odot}$ and ejects $4.77\:{\rm M_\odot}$ with $28\times
10^{51}\:{\rm ergs}$ (Table \ref{tab:98bw}). It is the most energetic
explosion in the most compact, lowest mass progenitor considered by
\citet{woo99}. Among their models for the supernova's explosion, they
find it gives the best fit to the observed bolometric and
spectroscopic light curves; however, significant discrepancies, such
as the fast observed rise in its luminosity,
remain.\footnote{\citet{iwa98} favor a more massive ($M_{\rm ej}\sim
10\:{\rm M_{\odot}}$) progenitor, with a correspondingly more
energetic (up to $\sim 5\times 10^{52}\:{\rm ergs}$) explosion,
giving similar values for $\ein$.}

\citet{woo99} used a nonrelativistic code with logarithmically smooth
zoning down to $10^{-6}\:{\rm M_{\odot}}$.  Estimating the energetics
of the radiation flash associated with shock break-out, they concluded
it was far too weak ($10^{42-46}$~erg) to power the GRB.
\citet{woo99} also estimated the amount of relativistic mass ejection,
by extrapolating their simulation to small masses with the shock
acceleration formula of \citet{gna85}, given here as equation
(\ref{gnatyk}). On this basis, they estimated $10^{44}-10^{45}\:{\rm
ergs}$ in kinetic energy of material with $\Gamma>3$ and
$10^{41}-10^{42} \:{\rm ergs}$ for $\Gamma>10$. However, as
trans-relativistic postshock acceleration (\S \ref{shacc}) was not
included, and as we have improved upon Gnatyk's formula in
\S\ref{S:theory}, we wish to re-examine the topic of relativistic
ejection from model CO6.

There are four methods we may employ to execute this
investigation. First, knowing the mass, radius, luminosity and
composition of the progenitor, we may use the analytical predictions
of \S \ref{S:whips} to give $E_k(\Gamma_f\beta_f>1)$. For this
progenitor, bound-free absorption dominates in an outer mass $M_{\rm
rel}$. Equation (\ref{ErelKramers}), which neglects radiation
pressure, predicts $M_{\rm rel} \simeq 0.99\times 10^{27}$ g with
kinetic energy $E_k(>1) \simeq 8.0 \times 10^{47}$~erg. In fact as
the luminosity of CO6 is about a quarter of the (electron-scattering)
Eddington limit, radiation pressure is not negligible, $n>3.25$ and
the above estimates are lower limits.

Second, if the outer density distribution of the progenitor were well
fit by a power-law density ramp (eq. [\ref{defineFrho1}]), then with
the fiducial explosion parameters of Table \ref{tab:98bw}, we could
evaluate $M_{\rm rel}$ and $E_k(>1)$ directly from equations
(\ref{MrelForEfficiency}) and (\ref{EkFromMrel3}). However, for model
CO6 no single pair of values for $n$ and $f_\rho$ accurately describe
the whole of the outer region. In particular the structure changes
around $\Mx\sim 1\times 10^{-4}\:{\rm M_\odot}$. The layers just
interior to this point are well described by $n\simeq9$ and
$f_\rho\simeq1.4$, while exterior we find $n\simeq 4$ and
$f_\rho\simeq 0.3$. To obtain the energy distribution as a function of
velocity, $E_k(>\Gamma\beta)$, we employ equation (\ref{EkGeneral})
with $A=0.736$, motivated by equations (\ref{Anr}) and (\ref{sigma}),
and the interior power-law fit ($n=9,f_\rho=1.4$). Internal variations
in the density structure effect the accuracy of this normalization of
$E_k$. Beyond the location where the structure changes to the $n=4$
case we extrapolate using equation (\ref{Ekscaling}). This ``broken
power-law'' estimate is shown by the {\it dashed} and {\it dotted}
lines in Figure \ref{fig:ke98bw}.
By this method we find $E_k(>1)\simeq 2.5\times 10^{48}$ erg and,
from equation (\ref{MrelForEfficiency}), $M_{\rm
rel}\simeq 3.0\times10^{27}\:{\rm g}$.

A third, more accurate, method 
is possible: since we have the full density distribution of the
progenitor model, we may directly apply equations (\ref{relshockf}),
(\ref{planar}), and (\ref{fsph}) to obtain the distribution of ejecta
at all velocities sufficiently fast for these equations to hold. This
technique accounts for interior density variations, which affect the
nonrelativistic shock propagation and thus the normalization of the
broken power-law estimate, which is based solely on the structure of
the outer layers.
Applying this more detailed approach is complicated by the fact that
the outermost zone of model CO6 is about $2\times 10^{27}$ g in
mass; this is the same order of magnitude as the estimates of $M_{\rm
rel}$ identified above. 
For accurate estimates, it is clearly necessary to continue the
sub-surface density distribution to much lower values. We therefore
extrapolate model CO6 using a power-law atmosphere with $n=4$, matched
to the outer edge of the original model.  After this extrapolation,
the smallest zone of the stellar model has a mass of $10^{-9}\:{\rm
M_{\odot}}$, several thousand times smaller than $M_{\rm rel}$; yet,
this zone is still optically thick to the high energy photons that
drive the outward shock \citep[a necessity for it to participate in
shock acceleration;][]{mat99}. We find that our results for the
mildly relativistic ejecta are not very sensitive to our choice of $n$,
nor to the exact value of the minimum zone mass.
Applying equations (\ref{relshockf}), (\ref{planar}), and (\ref{fsph})
to the augmented progenitor model, we find 
$M_{\rm rel} =3.9\times 10^{27}$ g and $E_k(>1)= 2.1\times 10^{48}$
erg. These values are within $\pm30\%$ of those obtained above 
by power-law fits. The full energy distribution is shown by
the central {\it long dashed} line in Figure \ref{fig:ke98bw}. The
upper and lower {\it long dashed} lines show the cases of
$\Ein=4.2\times 10^{52}$ and $1.9\times10^{52}\:{\rm ergs}$,
i.e. increasing and decreasing the fiducial explosion energy by a
factor of 1.5.

The final and most authoritative method to determine $M_{\rm rel}$ and
$E_k(>1)$ is by direct numerical simulation. Following \citet{woo99},
we simulate the explosion of CO6 by subtracting a remnant mass of
$1.78\:{\rm M_{\odot}}$ from the center and depositing $\Ein=2.8\times
10^{52}$ ergs in several pressurized, stationary zones; ignoring the
effects of gravity, we follow the subsequent behavior until the final
state of free expansion (for the relativistic ejecta, $r>9R$
suffices).  
We embed the progenitor in an optically thick stellar wind described
by the models of \cite{1994A&A...289..505S} \citep[see also][]{heg96},
considering mass loss rates of $10^{-6}$, $10^{-5}$ and $10^{-4}\:{\rm
M_{\odot}\:yr^{-1}}$. The wind velocities are relatively low near the
boundary of the star, increasing smoothly to achieve $90\%$ of the
terminal velocity, $v_w$, within $10-25~R$. \citet{heg96} set $v_w$
equal to the escape velocity from the stellar surface, which gives a
velocity of $3.6\times 10^{8}\:{\rm cm\:s^{-1}}$ for our
progenitor. We note this estimate of $v_w$ is uncertain and observed
velocities from WC stars are about a factor of two lower
\citep{koe95}. However, we find that the detailed structure of the 
stellar wind is of little consequence in our simulations of the
explosion. After the outermost ejecta have expanded to $9~R$ their
thermal energy has dropped to only a few percent of their kinetic and
postshock acceleration is essentially complete. At this point, only
$\sim 10^{-8}~\Msun$ has been affected by the reverse shock caused by
its interaction with the wind (see Figure
\ref{fig:displayejecta}). This justifies the approximation employed by
\cite{mat99}, and also in \S \ref{S:theory} above, that the production 
of high-velocity ejecta is unaffected by the existence of
circumstellar material. We shall argue in \S\S \ref{S:gammarays} and
\ref{S:radio} that GRB 980425 and the radio emission from SN 1998bw
were due to the circumstellar interaction; however, this occurs at
much larger radii, where the wind has reached its terminal velocity.

Our numerical simulation places $2.3\times 10^{48}$ ergs in $4.1\times
10^{27}$~g with $\Gamma_f\beta_f>1$, in excellent agreement with the
values predicted by our shock and postshock theory
(i.e., the third method listed above).
The energy distribution of the simulation is shown in Figure
\ref{fig:ke98bw} by the {\it solid} line. We conclude that our
equations describing the shock speed and the postshock acceleration of
trans-relativistic ejecta do an excellent job of predicting its energy
distribution with velocity -- except, of course, for material that has
been swept into the reverse shock ($\log_{10}(\Gamma_f \beta_f) >
0.6$). The upper and lower limits for the observed $\gamma$-ray energy
of GRB 980425, assuming it was associated with SN 1998bw, are shown by
the {\it hatched} area in Figure \ref{fig:ke98bw}. For our fiducial
model, we predict ejecta traveling faster than $\Gamma \beta
\gtrsim 1.35$, corresponding to a velocity $\beta \gtrsim 0.8$, have
$\Ek=E_\gamma$. Note that an explosion with twice our fiducial energy in a
$10\:{\rm M_{\odot}}$ progenitor, as considered by \citet{iwa98}, has the
same value of $\ein$, but results in $E_{\rm k}(>\Gamma\beta)$ twice
as large at any given velocity, assuming the same density structure.

Compared to \citet{woo99}, we find two to three orders of magnitude
more energy in ejecta traveling at mildly relativistic
($\Gamma_f\sim3$) velocities. This is principally because of our
inclusion of trans-relativistic postshock acceleration, which boosts
the final $\Gamma\beta$ of the ejecta by a factor of $\sim[2+(\Gamma_s
\beta_s)^{\sqrt{3}}]$. The steeply declining density gradient of the
progenitor, with $n=4$, results in $\Ek(>\Gamma_f\beta_f)\propto
(\Gamma_f\beta_f)^{-4.6}$ (eq. \ref{EkLimitNR}) in the nonrelativistic
limit - a very rapid falloff with velocity. Increasing the final
velocities of {\it all} the outer ejecta by even a modest factor
significantly increases the kinetic energy above a {\it particular}
$\Gamma_f\beta_f$. While we find enough energy in mildly relativistic
ejecta to account for the energetics of GRB 980425, the
ultrarelativistic ejecta are too puny to be of consequence. We now
examine if a mildly relativistic explosion can explain the observed
gamma-ray burst and radio supernova.

\subsubsection{Early Evolution - Gamma-Ray Burst}\label{S:gammarays}

We assume $\gamma$-rays are produced with efficiency $\epsilon_\gamma$
as the kinetic energy from relativistic ejecta is released on
collision with an effectively stationary circumstellar wind. For a
mass $m_{\rm ex}$ of ejecta traveling at a (mass-weighted) mean
Lorentz factor $\bar{\Gamma}\equiv 1+E_k(m_{\rm ex})/(m_{\rm ex}c^2)$,
about half of the initial kinetic energy $E_k(m_{\rm ex})$ will be
liberated after a mass $\sim m_{\rm ex}/\bar{\Gamma}$ of circumstellar
matter has been swept up \citep[e.g.,][]{pir99}. This estimate is
based on energy and momentum conservation in the wind frame, assuming
adiabatic interaction. Accounting for energy radiated away, 
leads to less wind material being required. For a wind of
constant velocity $v_w$ and mass loss rate $\dot{M}_w$, the mass
contained in a sphere of radius $r\gg R$ around the star is
$M_w=\dot{M}_w r/v_w$. Therefore, to liberate an amount of kinetic
energy $E_\gamma$ requires an interaction with the wind material out
to the radius
\begin{equation}\label{eq:r(Egamma)}
r(E_\gamma) \simeq \frac{E_\gamma
v_w}{\epsilon_\gamma \bar{\Gamma}(\bar{\Gamma}-1)\dot{M}_w c^2}, 
\end{equation}
where $\epsilon_\gamma$ cannot be much greater than a half.

Let us suppose that the shocked shell driven outward by the ejecta
mass $m_{\rm ex}$ travels with mean Lorentz factor $\Gamma_{\rm
shell}$; this is also the Lorentz factor of the shock that entrains
the wind, if it is fully radiative. Ignoring the fact that this shell
decelerates, the time for it to reach radius $r$ is $t^\prime \simeq
r/(\beta_{\rm shell}c)$ in the wind's frame. However, the observed
time of the burst, $t_{\rm obs}^\prime$\footnote{$t_{\rm obs}^\prime=t_{\rm obs}/(1+z)$, where
$t_{\rm obs}$ is the actual observed time at Earth and $z$ is the redshift of the burst.},
is shorter by a factor $1-\beta_{\rm shell}$ (still
ignoring deceleration), so that
\begin{equation} \label{eq:r(tobs)}
r(t_{\rm obs}^\prime) \simeq \frac{c t_{\rm obs}^\prime}{\beta_{\rm shell}^{-1}-1}
\simeq 2\Gamma_{\rm shell}^2 c t_{\rm obs}^\prime. 
\end{equation}
Here, the approximation $\beta_{\rm shell}^{-1}-1\simeq
1/(2\Gamma_{\rm shell}^2)$ overestimates $r(t_{\rm obs}^\prime)$ for 
trans-relativistic motions (for instance, by $37\%$ when $\Gamma_{\rm
shell}=1.7$), but this should roughly compensate for our neglect of the
shell's deceleration, which causes us to underestimate $r(t_{\rm obs}^\prime)$. 

What is the relationship between the Lorentz factor of the shell
($\Gamma_{\rm shell}(r)$), the mean Lorentz factor of the ejecta that
has struck the shell ($\bar{\Gamma}(r)$), and the Lorentz factor of
ejecta currently striking the shell ($\Gamma_f(r)$, say)? We must have
$\bar{\Gamma}(r)>\Gamma_f(r)$ because slower ejecta collide with the
shell at later times. In fact, $\Gamma_f(\bar{\Gamma})=\Gamma_f(E_k)$
is given by the ejecta's energy distribution, e.g., by equation
(\ref{EkGeneral}); for instance, we find that to have $\bar{\Gamma}=2$
requires $\Gamma_f \simeq 1.7$, for a variety of values of $n$.  Also,
$\Gamma_{\rm shell}<\Gamma_f$, because the ejecta must be able to
catch up with the shell. However, we expect that the latter quantities
differ by a relatively small amount.

The total isotropic energy of GRB 980425 in photons above 24 keV,
assuming association with SN 1998bw, is $E_\gamma =(8.1\pm1.0)
\times10^{47}\:{\rm ergs}$; its observed duration is $t_{\rm obs}\sim
35\:{\rm s}$ and this is approximately equal to $t_{\rm
obs}^{\prime}$, because the redshift is negligible.  Requiring that
$r(t_{\rm obs}^\prime) = r(E_\gamma)$ in equations
(\ref{eq:r(Egamma)}) and (\ref{eq:r(tobs)}) gives the following
estimate of the wind's mass loss rate:
\begin{eqnarray}\label{mdot}
\frac{\dot{M}_{-4}}{v_{w,8}} & \sim & \frac{3}
{(\bar{\Gamma}-1)(\bar{\Gamma}/{2})({\Gamma_{\rm shell}}/{1.7})^2}\nonumber
\\ &\times & \left(\frac{\epsilon_\gamma}{0.5}\right)^{-1}
\left(\frac{E_{\gamma}}{10^{48}\:{\rm erg}}\right)
\left(\frac{t_{\rm obs}^\prime}{35\:{\rm s}}\right)^{-1},~~~~~~
\end{eqnarray}
where $\dot{M}_{-4}=\dot{M}_w/10^{-4}\:{\rm M_{\odot}\:yr^{-1}}$ and
$v_{w,8}=v_w/10^8\:{\rm cm\:s^{-1}}$. 

In our fiducial model for SN 1998bw, we find 
ejecta with $\bar{\Gamma}\simeq 2$ and $\Gamma_f \simeq 1.7$ carry an
energy equal to $E_\gamma$. Ejecta with $\bar{\Gamma}\simeq 1.6$ and
$\Gamma_f \simeq 1.4$ contain a few $\times 10^{48}\:{\rm ergs}$. If
we take $\Gamma_{\rm shell}\sim \Gamma_f$ and $v_w\sim 10^8~{\rm
cm~s^{-1}}$, equation (\ref{mdot}) indicates that pre-supernova mass
loss rates of order a few$\times 10^{-4}\:{\rm M_{\odot}\:yr^{-1}}$
are required to explain the observed 35~s duration of GRB 980425. Note
that the radius of the circumstellar interaction is about $400~R$,
far outside the region in which the wind undergoes its acceleration
\citep{1994A&A...289..505S}.

Wolf-Rayet stars have mass loss rates typically in the range
$10^{-6}-10^{-4}\:{\rm M_{\odot}\:yr^{-1}}$ \citep{ham95}. However, it
is interesting that those in the the carbon-rich (WC) subclass are at
the high end of this range. \citet{koe95} present observations of 25
WC stars, with a mean mass loss rate of $\sim 6\times10^{-5}\:{\rm
M_{\odot}\:yr^{-1}}$ and about one order of magnitude dispersion.
To produce GRB 980425 from the mildly relativistic ejecta emerging from
SN 1998bw, we conclude that the circumstellar medium be dense but
within the range observed around Wolf-Rayet stars. 

While a dense circumstellar environment is necessary to liberate the
relativistic kinetic energy of the ejecta, it should be checked that
the $\gamma$-rays thus produced can escape this region to be observed
at Earth. For this we calculate the photospheric radius, $R_p$,
outside of which the $\gamma$-ray optical depth is $\tau=2/3$:
\begin{equation}
\label{tau}
\tau=\frac{2}{3}=f_{\rm kn}\kappa_{\rm es}\int^{\infty}_{R_p}\rho_w dr =
\frac{\dot{M}_w f_{\rm kn}\kappa_{\rm es}}{4 \pi v_w R_p}.
\end{equation}
Here, $\kappa_{\rm es}=0.40/\mu_e$~cm$^2$~g$^{-1}$, where
$\mu_e=2/(1+X_H)\simeq 2$ is the mean molecular weight per electron,
and $f_{\rm kn}$ is the relativistic (Klein-Nishina) correction to the
Thompson cross section; $f_{\rm kn} = 0.53,0.74,0.84$ for
$\gamma$-rays of energy $300,100,50\:{\rm keV}$. Thus, we have
\begin{equation}
\label{Rp}
R_p=1.5\times 10^{12} f_{\rm kn}\frac{\dot{M}_{-4}}{v_{w,8}} {~\rm cm},
\end{equation}
Although this photospheric radius ($\sim 100 R$) is much bigger than
the star, it is not larger than the interaction radius
$r(E_\gamma)=r(t_{\rm obs}^\prime)\sim 400 R$ identified above,
implying that the GRB will indeed be visible.
Additionally, equation (\ref{Rp}) predicts smaller photospheric radii
and earlier emergence for higher energy photons because of the energy
dependence of the Klein-Nishina opacity. For typical conditions inside
$R_p$ these higher energy photons are expected to be present
\citep{wea76}, and indeed the observed lightcurves of GRB 980425 peak
a few seconds earlier for higher ($>40$ keV) energies than for lower
($<26$ keV) energies \citep{2000ApJ...536..778P}.

As the emphasis of this paper is on the energetics of relativistic
mass ejection, we shall not attempt to calculate in any detail the
spectrum of the observed outburst; however, note that the typical
energies of these photons will depend on the mechanism of their
emission. 
Non-thermal processes include inverse Compton upscattering of lower
energy photons, or synchrotron emission by electrons produced in the
shock front. As the latter is the canonical mechanism for the emission
from GRBs, we give a simple estimate of the typical energies of these
photons. Consider a shock of Lorentz factor $\Gamma_{\rm sh}$
propagating into the wind (density $\rho_w = \dot{M}_w/(4\pi r^2
v_w)$). The postshock Lorentz factor $\Gamma_2$ is related to
$\Gamma_{\rm sh}$ by equation (\ref{jump}). The postshock energy
density \citep{bla76}, minus the rest energy of the postshock fluid,
is the energy density associated with the postshock pressure:
\begin{equation} \label{eq:e2}
e_2 - \rho_2 c^2 = (4\Gamma_2+3)(\Gamma_2-1)\rho_w c^2. 
\end{equation} 
Following \cite{pir99}, we assume that electrons and magnetic fields
account for fractions $\epsilon_e$ and $\epsilon_b$ of this amount,
respectively. The typical random Lorentz factor of an electron
in the postshock flow is then
\begin{equation} \label{gammabar_e}
\bar{\gamma}_{e} = \frac{\mu_e M_p}{M_e} \epsilon_e (\Gamma_2-1).  
\end{equation}
Approximating $\Gamma_{\rm shell}\simeq \Gamma_2$, and evaluating the
photon energy of synchrotron radiation by a typical electron at the
radius $r(t_{\rm obs}^\prime)$ given in equation (\ref{eq:r(tobs)}), we find
\begin{eqnarray}\label{synch}
h\nu_{\rm syn} &\simeq& 6.3 \mu_e^2 \Gamma_{\rm shell}
(\Gamma_{\rm shell}-1)^{1/2} 
(4\Gamma_{\rm shell}+3)^{1/2}
\nonumber \\ & \times &\epsilon_e^2 \epsilon_b^{1/2}
\left(\frac{\dot{M}_{-4}}{v_{w,8}}\right)^{1/2} 
\left(\frac{t_{\rm obs}^\prime}{35~{\rm s}}\right)^{-1}
 ~{\rm keV}.~~~~~
\end{eqnarray}
With $\mu_e\simeq 2$, this gives $h\nu_{\rm syn} \simeq (8, 14,
29, 73)\times (\epsilon_e/0.5)^2 (\epsilon_b/0.5)^{1/2} (35~{\rm s}/t_{\rm
obs}^\prime)\dot{M}_{-4}/v_{w,8}$~keV, for $\Gamma_{\rm shell} = (1.25, 1.5,
2, 3)$. The synchrotron model is thus roughly consistent with the photon
energies observed in the burst. It suggests that the softening of
these photons over time may have been caused by the shell's
deceleration combined with the declining pre-shock density.

The earlier arrival of harder photons thus occurs naturally in both
the optically thick regime, because $R_p$ is a decreasing function of
$h\nu$, and in the optically thin regime, because of the decline in
$h\nu_{\rm syn}$ as the shell expands. 

\subsubsection{Late Evolution - Radio Supernova}\label{S:radio}

Radio observations of SN 1998bw have been used to infer ejecta
expansion velocities, assuming the radio emission originates in a
shell just behind the forward shock as it propagates into the stellar
wind. \citet{kul98} argued the lack of strong variability at low
frequencies implies the shock reached a size $\gtrsim 10^{16}\:{\rm
cm}$, set by the refractive scintillation scale of the Galactic ISM,
by the time of the radio light curve peak at $\sim 12$ days. This
implies $\bar{\beta}_s \gtrsim 0.3$ averaged over this period. Furthermore,
\citet{kul98} pointed out that two constraints, one on the brightness 
temperature limited by inverse Compton scattering and the other on the
total energy of a synchrotron emitting source, require relativistic
($\Gamma \beta \sim 1.6 - 2$) shock expansion velocities up to 60 days
after the start of the supernova explosion.

\citet{wax99} and \citet{li99} have presented additional models of the
radio emission. \citet{wax99} considered a mono-energetic,
approximately thermal, electron energy distribution, to derive shock
expansion with $\beta_s\sim 0.3$ at $t=12$ days. They argued the radio
emission does not imply the presence of a highly relativistic blast
wave and that this weakens the link between SN 1998bw and GRB
980425. Their model required only a weak magnetic field, far from
equipartition, and a high density ($\sim 10^4 - 10^5$) of emitting
electrons. \citet{li99} modeled a power law electron energy
distribution and accounted for effects of relativistic expansion of
the synchrotron source. They derived a range of values of the total
energy, equipartition fractions of electrons, $\epsilon_e$, and
magnetic fields, $\epsilon_b$, and the ambient wind density that were
compatible with the radio observations. They considered a specific,
arbitrary model with $\epsilon_b=10^{-6}$ and a dense
($\dot{M}_{-4}/v_{w,8}=0.6$) wind, which gave $\Gamma_s
\beta_s\sim0.75$ after $12$ days. 

How do the above velocities compare to those predicted from our model?
The circumstellar interaction becomes nonrelativistic after a wind
mass $\sim M_{\rm rel}$ has been swept up; this happens at a radius
$r\gtrsim 10^{14} v_{w,8}/\dot{M}_{-4}$~cm, or after about one hour has
elapsed since the burst. The shock is therefore non- or mildly
relativistic ($\Gamma_{\rm shell} \simeq 1$) at the time of the radio
observations. We assume the forward, radiating shock expands according
to $r_s\propto t^{\eta}$, so that $\beta_s c = \eta r/t$ where
$0<\eta<1$.  The ejecta colliding with the shocked shell are in free
expansion ($\beta_f c = r/t$), so $\beta_s = \eta \beta_f$. The
nonrelativistic ejecta are distributed as $\rho_f \propto
\beta_f^{l_{\rho 2}}$, where $l_{\rho 2} \simeq -(8.3+5.3/n)$
\citep[e.g.,][]{mat99}. Because $l_{\rho 2}<-5$, the self-similar
solution of \cite{1982ApJ...258..790C} and \cite{nad85} applies:
there, the ratio of masses of shocked ejecta and
ambient material is a constant. Fitting the data in Chevalier's table,
we find 
\begin{equation}\label{Mratio}
\frac{\Mx}{M_w} = \frac{-l_{\rho 2}}{2.7} -1.8, 
\end{equation}
within four percent. 
The expansion parameter $\eta$ is determined by the requirement that
the ratio $\Mx/M_{w}$ be constant: 
\begin{equation} \label{eta}
\eta = \frac{n+1}{1.2n+1}. 
\end{equation}
Note that the deceleration is very slow ($\eta\simeq 1$);  this will
be crucial to the persistence of rapid expansion into the epoch of the
radio observations. 

Requiring that the ambient mass within $r$ and the ejecta mass
traveling faster than $\beta_f = \beta_s/\eta$ satisfy equation
(\ref{Mratio}), and using equations (\ref{nonrelshock}),
(\ref{planar}), (\ref{spherical}), and (\ref{defineFrho1}) to specify
$M_{\rm ej}(>\beta_f)$, we find that for our fit ($n = 4$,
$f_\rho = 0.3$) to the outer envelope,
\begin{eqnarray} \label{latetime}
\beta_{\rm shell} \simeq 0.42
\left( \frac{\Ein}{2.8\times10^{52}\:{\rm
  ergs}}\right)^{0.43} \left(\frac{v_{w,8}}{\dot{M}_{-4}
}\right)^{0.13}
\nonumber\\ 
\times  \left(\frac{\Mej}{4.77\:{\rm M_{\odot}}}\right)^{-0.30}
  \left(\frac{t}{1\:{\rm day}}\right)^{-0.13}. 
\end{eqnarray}
After twelve days $\beta_s\simeq0.30$; the velocity declines only
slowly because the amount of ejecta mass traveling faster than a given
$\beta_f$ increases steeply as $\beta_f$ decreases. The average
velocity is faster by only $15\%$, due to the slowness of the
deceleration, so the mean rate of expansion is $\bar\beta_s = 0.35$ at
twelve days. This velocity satisfies the constraints from the radio
scintillation measurements, and is consistent with synchrotron
emission models, given their uncertainties, e.g. in the assumed
electron distribution. Including relativistic effects would not much
alter our result because these have two counteracting effects: on the
one hand, they increase the final velocities of ejecta, thus raising
the velocity relative to the above equation; but on the other, they
lead to a smaller swept-up mass being required to slow down the
ejecta, thus reducing the velocity.

We conclude that an energetic spherical explosion of a few $\times
10^{52}\:{\rm ergs}$ in a compact, relatively low mass carbon-oxygen
core of a massive star is capable of producing enough relativistic
ejecta to account for the energetics of GRB 980425. The results of
\citet{woo99} favor this type of progenitor from fits to the light 
curve and spectra, although the explosion energy we require is at the
high end of the range they considered.  The timescale of the burst
requires a relatively dense stellar wind around the CO core, with
implied mass loss rates $\sim {\rm few}\times 10^{-4}\:{\rm
M_{\odot}\:yr^{-1}}$. Gamma-ray photons are able to escape from this
wind to be observed at Earth. Note that opacity due to photon-photon
interactions (see \S\ref{S:C-GRBs}) does not constrain the minimum
Lorentz factor of the ejecta because the observed GRB spectrum is very
soft. These typical wind densities also allow $\gamma$-rays of the
correct energy to be produced by synchrotron emission behind the
mildly relativistic forward shock. The shock velocity into this dense
wind is fast enough to explain the radio properties of SN 1998bw days
after the GRB. This simple spherical model is thus able to account for
the observed properties of GRB 980425, if indeed it was associated
with SN 1998bw.

\subsection{Hypernovae and Cosmological Gamma-Ray Bursts}\label{S:C-GRBs}

\begin{deluxetable}{cccccccc} 
\tablecaption{Conservative lower limits on $\Gamma$ in the
external shock model for GRBs with afterglows
\label{tab:CGRBs}}
\tablehead{
\colhead{\scriptsize GRB} & \colhead{\scriptsize Redshift, z} & \colhead{\scriptsize Isotropic $E_\gamma$}
& \colhead{\scriptsize Duration (s)}
&\colhead{\scriptsize Min. $\bar\Gamma$ (eq. \ref{GammaOpticallyThin2})}
&\colhead{\scriptsize Min. $\bar\Gamma$ (eq. \ref{photonphoton}; $\alpha=2$)}
&\colhead{\scriptsize Refs.\tablenotemark{a}}  \\
             &                     &\colhead{\scriptsize ($10^{52}$~erg)} 
&\colhead{\scriptsize $t_{\rm obs}=t_{\rm obs}^\prime(1+z)$}                          & 
 \scriptsize [external opacity]  & \scriptsize [$\gamma-\gamma$ opacity]
& } 
\startdata
970228 & 0.695 &0.5 & 80  & 9  &9 & 1, 2, 3  \\
970508 & 0.835 &0.8 & 15  & 17 &14 &3, 4  \\
971214 & 3.418 &30  & 25  & 36 &25 &5, 6 \\
980703 & 0.966 &10  & 100 & 14 &12& 7, 8\\
990123 & 1.600 &300 & 100 & 28 &20& 9, 10\\
990510 & 1.619 &30  & 80  & 20 &15& 3, 11 \\
991208 & 0.706 &13  & 60  & 17 &14& 12, 13 \\
000301C&  2.03 &0.2 & 10  & 19 &15& 14, 15 \\
000418&  1.119 &4.9 & 30  & 15 &13& 16, 17\\
000926&  2.066 &26 & 25  & 31 &22& 18, 19\\
\enddata
\tablenotetext{a}{\scriptsize Table is based on data compiled by J. Greiner and S. Hempelmann 
at www.aip.de/$\sim$jcg/grbrsh.html. (1) \citet{cos97}; (2) \citet{djo99}; (3)
\citet{1999ApJ...523L.121H}; (4) \citet{cos97b}; (5) \citet{hei97}; (6)
\citet{mes99};
(7) from http://gammaray.msfc.nasa.gov/batse/grb/lightcurve/; (8)
\citet{1998ApJ...508L..21B}; (9) \citet{fer99}; (10) \citet{kul99}; (11)
\citet{dad99}; (12) \citet{hur99}; (13) \citet{djo99b}; (14)
\citet{smi00}; (15) \citet{cas00}; (16) \citet{hur00}; (17) \citet{blo00}; 
(18) \citet{hur00b}; (19) \citet{fyn00}.}
\end{deluxetable}

\begin{figure}
\epsscale{1.0}
\plotone{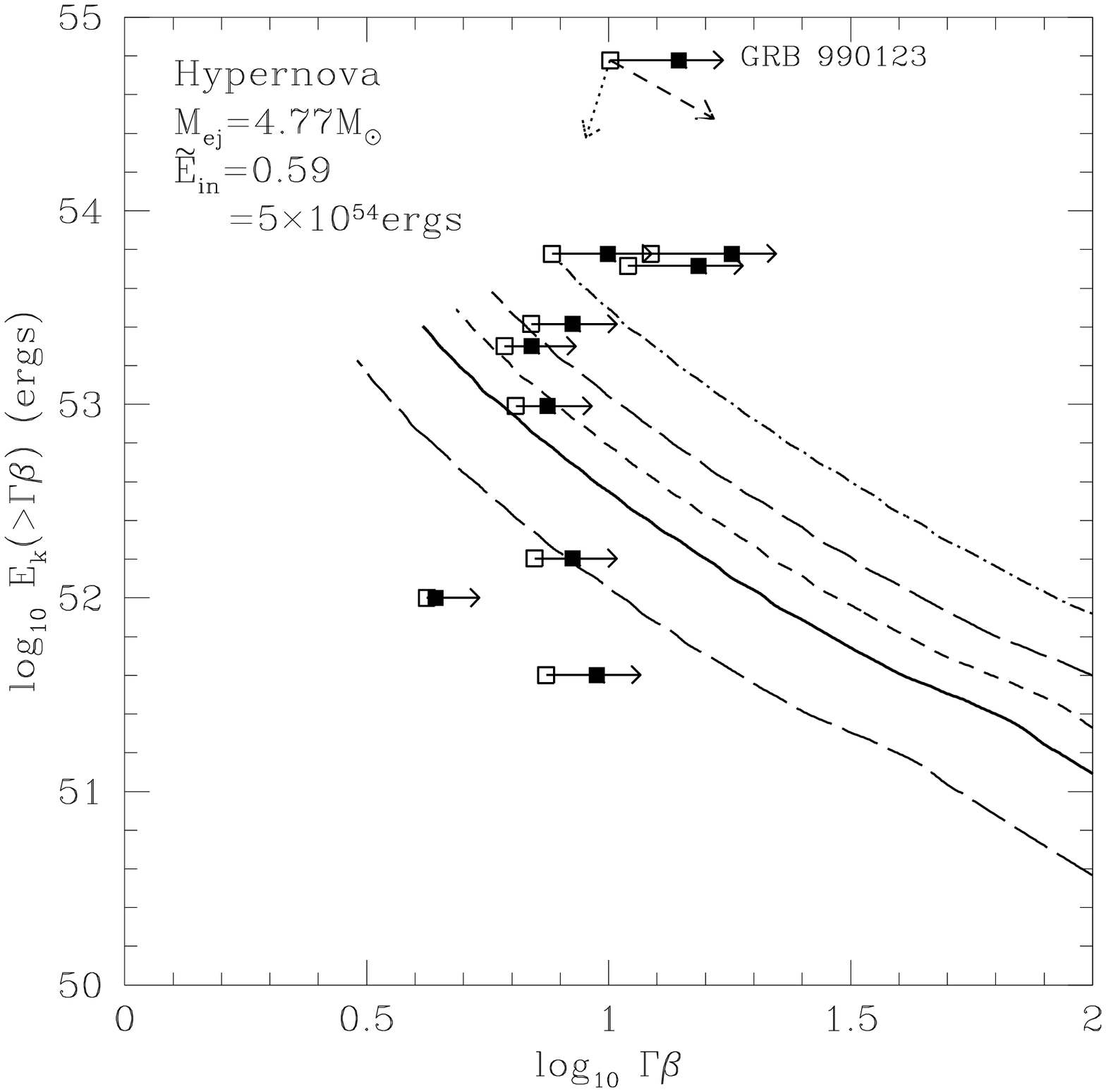}
\caption{
Cumulative kinetic energy distributions of ejecta from extreme
hypernova explosions, modeled in our SN 1998bw progenitor.
Predictions are based on the analytic approximations of
\S\ref{S:theory}. The fiducial model is shown by the {\it solid}
line. The upper and lower {\it long-dashed} lines show the effect of
increasing and reducing $\Ein$ by a factor of 1.5,
respectively. Uncertainties in the spherical correction factor for
postshock acceleration lead to $\sim 25\%$ uncertainty in the final
energy distribution.  The shock velocity normalization coefficient
employed in these models is $A=0.736$. However, there is some evidence
from simulation that this value may increase for $\ein\sim{\cal
O}(1)$. The fiducial model with $A=0.8$ results in the {\it dashed}
line. Combining these two factors, with the increased value of $\Ein$
results in a maximal model ({\it dot-dashed} line). The {\it squares}
show the required isotropic energies $E_\gamma/\epsilon_\gamma$, with
$\epsilon_\gamma=0.5$, of the cosmological GRBs with detected optical
afterglows, plotted at the minimum values of $\Gamma_f\beta_f$
implied by external ({\it solid}) and $\gamma -
\gamma$ ({\it open}) opacity constraints (Table
\ref{tab:CGRBs}). Note that the minimum $\Gamma_f \simeq \bar{\Gamma}/2$. The 
{\it dotted arrow} shows the effect on the $\gamma - \gamma$ opacity
constraint for GRB 990123 of adopting a smaller $E_\gamma = 1.2\times
10^{54}\:{\rm ergs}$ \citep[e.g.][]{bri99}, consistent within
cosmological uncertainties, while the {\it dashed arrow} shows the
effect of applying the constraint to the first major pulse of the
lightcurve with FWHM $\sim 10$ s and energy $\sim 1/2$ of the total.
\label{fig:cgrb}}
\end{figure}

In \S \ref{S:98bw} we found that GRB~980425 and the radio afterglow of
SN~1998bw can plausibly be attributed to the collision of mildly
relativistic ejecta with a circumstellar wind. An interaction of this
type has often been considered as an important possibility for GRB
afterglows
\citep{1998ApJ...496L...1R,1998ApJ...503..314P,2000ApJ...535L..33S}.
External shock models have also been proposed
\citep{1993ApJ...405..278M} for the emission from cosmological
GRBs; however, these have been criticized
\citep[e.g.,][]{sar97,pir99} on the basis that they cannot reproduce the
substructure observed in many GRBs. This constraint does not apply to
GRB~980425, whose profile was smooth, nor to the few other bursts
\citep{1999ApJ...518..901N} that share this property. Could
cosmological bursts with smooth $\gamma$-ray light curves be
produced by spherical explosions of over-energetic supernovae
\citep[``hypernovae'';][]{woo93,pac98}, in the same manner that 
GRB~980425 was produced by SN~1998bw? 

The cosmological GRBs with known redshifts require explosion energies
of $10^{51.5-54.5}$~erg, i.e., $(0.002 - 2)\times \Msun c^2$ to be
emitted as $\gamma$-rays, under the assumption of isotropic
emission. By extracting energy efficiently from the formation of a
massive black hole at its center,
a hypernova may be capable of such energies.  (Note that the stellar-mass Galactic
black holes are clustered around $7~\Msun$;
\citealt{1998ApJ...499..367B}). However, in a spherical explosion of the
type advocated for SN~1998bw, much of the energy is locked up in
ejecta moving near the mean Lorentz factor $(1+\ein)$. It is well
known that the time scales and energies of GRBs require significant
Lorentz factors to be optically thin; this would imply an even larger
energy in low-velocity ejecta, if the necessary value of $\Gamma$
significantly exceeds $1+\ein$.

There are two potential sources of opacity that can shade GRB
emission, and hence two constraints on the Lorentz factor.  The first
of these is absorption of the observed photons in pair-producing
interactions with other photons
\citep{1991ApJ...373..277K,1997ApJ...491..663B}; this leads to the
``compactness problem'' for GRBs.  The cross section is highest when
the center-of-momentum energy is somewhat above $2m_e c^2$, so one
must typically assume the number of photons varies as $N_\gamma(>\nu)
\propto \nu^{-\alpha}$, where $\alpha > 1$, to extrapolate from the
observed photons to those that might absorb them (those with energies
$\sim \Gamma^2 (m_e c^2)^2/(h\nu)$, typically not observed). Our
criterion for the fireball to be transparent is that
$\tau_{\gamma\gamma}\leq 2/3$ for the observed photons. This requires that the
Lorentz factor satisfy \citep{pir99}
\begin{equation}\label{photonphoton}
\Gamma^{4+2\alpha} \gtrsim 1.1\times 10^8 \left(\frac{ E_{\gamma}}{10^{52}~\rm
erg}\right)\left(\frac{ t_{\rm obs}^\prime}{35~\rm s}\right)^{-2},
\end{equation}
assuming the fraction of observed photon pairs with sufficient 
energy to pair produce is unity (an over estimate). Note, $t_{\rm
obs}^\prime=t_{\rm obs}/(1+z)$, where $t_{\rm obs}$ is the
observed duration at Earth and $z$ is the redshift of the GRB. Direct
observations of the high energy (1 - 20 MeV) photons from GRB 990123
imply $\alpha\simeq 2$ \citep{bri99}, and we adopt this value, as it
is typical of many bursts. ``No High Energy'' bursts, such as GRB
980425, with $\alpha>4.5$ between 100 and 300 keV
\citep{pen97}, are not severely constrained by this opacity. There is
some evidence for harder spectra ($\alpha\sim 1-1.5$), up to
100 MeV - GeV energies in a few bursts where observations are
available (over the same time periods as the keV burst)
\citep{hur94,sch95,din97}, and such hard slopes would increase the
minimum Lorentz factor.
The value of $\Gamma$ in equation (\ref{photonphoton}) ought to
represent a frame in which the photon distribution is isotropic, i.e.,
the frame of the cooling postshock fluid; for simplicity, we shall
identify this with the mean (mass-weighted) Lorentz factor of the
relevant portion of the ejecta, $\bar\Gamma$.

In an external-shock model, the surrounding material whose presence is
necessary to liberate the burst energy inevitably introduces a second
source of opacity. Adopting a wind model ($\rho\propto r^{-2}$) for
this material, and requiring its optical depth be less than $2/3$
(i.e., requiring $r(E_\gamma)=r(t_{\rm obs}^\prime)>R_p$ in equations
[\ref{eq:r(Egamma)}], [\ref{eq:r(tobs)}], and [\ref{Rp}]), we find
\begin{equation} \label{GammaOpticallyThin}
\bar{\Gamma} (\bar{\Gamma}-1) \Gamma_{\rm shell}^4 > \frac{3E_\gamma
f_{\rm kn} \kappa_{\rm es} }{32 \pi \epsilon_\gamma t_{\rm obs}^{\prime 2} c^4 }. 
\end{equation} 
Note, $\Gamma_f \simeq \Gamma_{\rm shell} \simeq \bar\Gamma(>\Gamma_f)
/ 2$, for $n = 4$ and $\Gamma_f\gg 1$, (equation
\ref{EkIntermediateR}), and so 
\begin{equation}  \label{GammaOpticallyThin2}
\bar{\Gamma} \gtrsim 11 
\left(\frac{ E_{\gamma}}{10^{52}~\rm erg}\frac{0.5}{\epsilon_\gamma}\right)^{1/6} 
\left(\frac{ t_{\rm obs}^\prime}{35~\rm s}\right)^{-1/3}, 
\end{equation} 
as long as the implied $\bar{\Gamma}\gg 1$.

Often, models of GRBs or their afterglows assume a uniform rather than
a wind ambient medium. In this case, the opacity of the swept-up gas
is not at issue, if the burst is not in an obscured region. In an 
external shock model, one may constrain $\bar \Gamma$ by assuming a
hydrogen number density $n_{\rm H}$ and requiring that sufficient mass be
swept up to thermalize the kinetic energy of the ejecta: 
\begin{equation} \label{uniformconstraint}
\bar \Gamma \simeq 88 
\left(\frac{E_\gamma}{10^{52}\:{\rm erg}}\frac{0.5}{\epsilon_\gamma}\right)^{1/8} 
\left(\frac{t_{\rm obs}^\prime}{35 {\rm s}}\right)^{-3/8} n_{\rm H}^{-1/8}.
\end{equation}
This constraint on $\bar \Gamma$ is less severe in more dense regions,
which is why the opacity constraint is the more relevant one for wind
interaction models. 

In an external-shock model, $t_{\rm obs}^\prime$ is the total duration of the
burst, and significant substructure is difficult to explain \citep{sar97};
in an internal-shock model, $t_{\rm obs}^\prime$ is the duration of an
individual pulse. In either case, the total duration sets a lower
limit for the Lorentz factor: we give these lower limits in Table
\ref{tab:CGRBs}, for ten bursts with known redshifts. Like GRB~980425, 
most of these bursts show little fine (sub-second) structure in their $\gamma$-ray
light curves. The profiles are quite simple with most of
the energy contained in one or two pulses lasting $\gtrsim 10\:{\rm
s}$. Note that some substructure may be produced by the
interaction of an external shock with an inhomogeneous medium
\citep{der99,fen99}. Instabilities in accelerating shocks \citep{1994ApJ...435..815L} 
may also create variability. Strong density jumps in the progenitor
may lead to variable energy release from the postshock flow as
different layers crash into the decelerating region. Structure was
observed in the radio emission from SN 1998bw
\citep{kul98,1999A&AS..138..467W}, which is indubitably an external
shock; the same can be said for fluctuations in other radio supernova
light curves \citep[e.g., 1979c;][]{wei00}, which are attributed to
variations in the ambient density.

To gauge how difficult it is for a spherical hypernova to satisfy the
above constraints on energy and velocity of relativistic ejecta, we
adopt an extreme explosion of $\Ein = 5\times 10^{54}~\rm erg$, perhaps
liberated during the formation of a $10\:{\rm M_{\odot}}$ black
hole. We note that models for magnetic extraction of this energy on
timescales short compared to the break-out time may require field
strengths in excess of $10^{15}\:\rm {G}$
\citep{mac86}. For lack of a better model, we surround the nascent
black hole with the $\sim 5\:{\rm M_{\odot}}$ envelope of model CO6
(\S \ref{S:98bw}). This gives $\ein\simeq0.59$, placing the explosion
in a regime untested by our simulations. Regardless, we employ our
analytic approximations to investigate the expected energy
distribution of the hypernova's ejecta (Figure \ref{fig:cgrb}).

We compare these predictions with the required ejecta energies and
minimum Lorentz factors from opacity constraints of GRBs with observed
optical afterglows, listed in Table
\ref{tab:CGRBs}. Note that $\Gamma_f \simeq \bar\Gamma(>\Gamma_f) / 2$,
for $n = 4$ and $\Gamma_f\gg 1$, (equation \ref{EkIntermediateR}). We
require $E_k(>\bar\Gamma/2) = E_\gamma/\epsilon_\gamma$. There is a
wide range of observed energies, yet quite similar minimum Lorentz
factors are implied from the photon-photon and external opacity
constraints, spanning a range from about 10 to 40. These values are
lower than those commonly quoted in the literature
\citep[e.g.][]{lit01}, because the bursts in question have relatively
smooth time profiles, and because our adoption of the external-shock
hypothesis dictates that the age of a burst, rather than the timescale
of its substructure, be used to constrain the Lorentz factor. Our
extreme hypernova models can satisfy these requirements in three, and
potentially six (given the uncertainties), of ten cases. However, the
most energetic bursts require preferential viewing of asymmetric
explosions (\S\ref{S:aspherical}). Since the most luminous GRBs do not
differ from typical GRBs in any significant manner except luminosity,
it is therefore likely that asymmetric explosions are required for all
cosmological GRBs. Nonetheless, these results demonstrate the
potential importance of the trans-relativistic shock acceleration
mechanism for GRB engines.

Afterglows potentially pose a difficult constraint on spherical models
invoking shock acceleration and external emission, as there must be
much more energy in slower ejecta than implied by the burst. We have
found $E_k(>\Gamma_f\beta_f)\propto 1/(\Gamma_f\beta_f)$ (equation \ref{EkLimitXR}),
approximately, for values of $\Gamma_f\beta_f$ significantly above $1+\ein$;
the energy rises even more sharply for low values of $\Gamma_f\beta_f$
(Figure \ref{fig:kean}). As the emitting region is continuously ``refreshed''
as it decelerates by this increasing source of energy
\citep{1998ApJ...496L...1R,2000ApJ...535L..33S}, afterglows should
evince these even higher energies. This is especially true if the
emitting shell makes a transition from radiative to adiabatic dynamics
between the burst and its afterglow.  \cite{fre99} argue that
afterglow observations indicate an energy comparable (within a factor
of three) to the energy emitted in $\gamma$-rays, despite an assumed 
deceleration from $\Gamma\gtrsim 100$ to $\Gamma\simeq 10$. Moreover,
the emitted $\gamma$-ray energy places a lower limit on the energy
content (per solid angle) of the emitting gas in its early phase. 

We use the results of \citet{2000ApJ...535L..33S} to evaluate the
severity of such afterglow constraints. Assuming a wind ambient medium 
$\rho\propto r^{-2}$ and ultrarelativistic ejecta from an
envelope with polytropic index $n$, equation (\ref{EkIntermediateR})
and equation (3) of \citeauthor{2000ApJ...535L..33S} gives a swept-up
kinetic energy that varies as 
\begin{equation}\label{EshellWindShock}
E_k\propto t_{\rm obs}^{\prime(2.7+n)/(2.7+7.7n)}, 
\end{equation}
i.e., $E_k\propto t_{\rm obs}^{\prime(0.2, 0.17, 0.13)}$ for $n=(4, 7,
\infty)$. The increase in kinetic energy from 35 seconds to ten days
is thus a factor of $\sim (7.6, 5.6, 3.7)$ in a model involving a
hypernova and its wind, for these three values of $n$. Considering the
uncertainties in the \citeauthor{fre99} results, we conclude that it
cannot yet be argued that afterglow calorimetry rules out
approximately spherical hypernovae as the origins of some cosmological
GRBs, especially those with relatively low $E_\gamma$ for which
\citeauthor{fre99} find the afterglow to be about three times more
energetic than the GRB.

All of the above constraints are somewhat assuaged if the progenitor
or its explosion are asymmetric. As discussed in \S
\ref{S:aspherical}, an assessment of the implications of asymmetrical
explosions requires that multidimensional simulations be interpreted
using our equation (\ref{eq:asphYp}). Under the admittedly
simplistic sector approximation, we found that relatively small
variations in the energy per unit mass of the ejecta can have
substantial effects on the energy of the resulting relativistic
ejecta.  More accurately, given a shock intensity $p$ below a given
column density on some patch of the stellar surface, the kinetic
energy in that direction (above some chosen velocity) varies as
$p^{5.35\gammap}$. For $n=(4,7)$, the shock intensity $p$ need only
be enhanced a factor $(1.99, 2.12)$ at some reference column to
increase the kinetic energy in that direction by a factor of one
hundred, at all velocities for which the ejecta experienced an
accelerating shock. This enhancement would be enough to bring even GRB
990123 within the reach of hypernova models. We eagerly await
simulations that will evaluate whether this degree of asymmetry is
realistic.

In summary, while our ad hoc and extreme spherical hypernova model is consistent
with many of the GRBs with observed afterglows, it appears likely that
the most energetic bursts require asymmetric progenitors or energy
injection. Given this possibility, trans-relativistic blast waves from
hypernovae may in principle account for the energetics and ejecta
velocities of cosmological bursts, though it is too early to decide if
this mechanism is primarily responsible for their origin. The
association of these cosmic GRBs with the stellar population of their
host galaxies \citep{blo00b}, and more tentatively with star forming
regions \citep{kul00, djo00, gal00b}, argues in favor of a connection
with the deaths of massive stars and against scenarios involving
neutron star mergers. The difficulty of avoiding baryon loading while
trying to escape the stellar progenitor, encountered even in highly
asymmetric jet models \citep{alo00}, is overcome in the hydrodynamic
shock acceleration mechanism. The efficiency of $\gamma$-ray
production is higher for a blast wave producing these photons from
both external and internal shocks, compared to one that radiates
solely via internal shocks
\citep{kum99}.  While external shocks are inefficient at producing
substructure \citep{sar97}, the lightcurves of the bursts with
observed afterglows are relatively smooth, with little power in
sub-second features. It remains to be demonstrated if our model can
account for moderate variability, perhaps via the interaction with
circumstellar inhomogeneities \citep{der99,fen99} or via the
production of postshock velocity perturbations by the growth of
instabilities in the accelerating shock \citep{1994ApJ...435..815L} or
by the presence of steep density shelves in the progenitor's
atmosphere.  Further investigation is also required to determine
whether a model with monotonically decelerating forward and reverse
shocks produces spectra \citep[see][]{2000ApJ...535L..33S} that
resemble those of cosmological GRBs as well as their afterglows.

\subsection{Ejecta from Compact Objects}\label{S:WDNS}

\begin{figure}
\epsscale{1.0}
\plotone{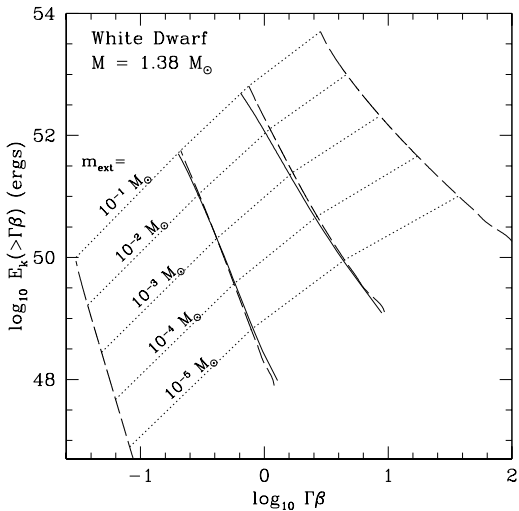}
\caption{Kinetic energy distribution with velocity for the outer 
$0.1\:{\rm M_{\odot}}$ of ejecta from white dwarf explosions.
Numerical simulation results are shown by the {\it solid} lines, and
analytic predictions from \S\ref{S:theory}, integrated over the
progenitor density distribution, by the {\it long dashed} lines. The
agreement is excellent for the outermost ejecta, but, as expected,
deviates towards the interior, where approximations of the analytic
theory break down. Models of AIC \citep{fry99} tend to place
$\sim10^{50}\:{\rm ergs}$ in $0.01-0.1\:{\rm M_{\odot}}$.
\label{fig:kewd}}
\end{figure}

\subsubsection{White Dwarfs}\label{S:WDs}
Accreting white dwarfs near the Chandrasekhar mass face two quite
distinct fates. Type Ia supernovae involve the release of $\sim
10^{51}\:{\rm ergs}$ in $1.4~\Msun$ ($\ein \sim 4\times 10^{-4}$), in
either a deflagration or a detonation
\citep{woo86}. Approximating a detonation as an adiabatic point
explosion, equation (\ref{EkGeneral}) with $f_\rho=0.3$,
$n=1.9$, and $A=0.705$ (values derived from a model provided by
Lee Lindblom) gives $\Ek(\Gamma_f>3,10)= (35, 1.1)\times 10^{39}\:{\rm
ergs}$, respectively \citep[see also][who presented an approximate
treatment of shock and postshock acceleration in WDs]{ber96}. These
energies are too small to produce observable GRBs.

Another possible outcome, if the central density is high enough to
allow electron capture prior to carbon or neon ignition, is
accretion-induced collapse (AIC) from white dwarf to neutron star
\citep[e.g.,][and references therein]{woo92,fry99}. Such events
resemble the core-collapse supernovae of massive stars, but lack the
ram pressure and long timescale introduced by an extended infalling
envelope. \cite{goo87} suggested that neutrino annihilations above the
neutrinosphere of a newly-formed neutron star could vent $\sim 0.3\%$
of the neutrino luminosity into the nearby volume; if this region were
baryon-free, a relativistic wind of energy $\sim 10^{50}$ erg would
result, which could drive a GRB \citep{dar92}. However, the
neutrinosphere underlies a massive layer of baryons \citep{woo92}, so
there results instead a neutrino-driven wind of velocity $\sim c/3$.
This may either expel an exterior envelope, or expand freely if there
is no envelope left \citep{fry99}. There is not sufficient energy in
such winds to power the cosmological GRBs listed in Table
\ref{tab:CGRBs}, and moreover they cannot satisfy the Lorentz factor 
constraint, (\ref{photonphoton}) or (\ref{GammaOpticallyThin2}). They
could only produce fast ejecta through shock acceleration; but, since
the mean velocity of the ejecta cannot exceed $c/3$ (\S
\ref{S:CharacteristicVelocities}), only a small fraction of the energy
is available at appreciable $\Gamma_f$.

In addition to the energy injection from this wind, there is also a
prompt shock from the core collapse \citep[which stalls;][]{fry99},
and ``delayed'' heating by neutrino absorption behind the stalled
shock; the latter drives ejection in the model of
\citeauthor{fry99}. This simulation has $\sim 10^{50}$ ergs
deposited in $\sim 0.1~\Msun$, for a mean expansion velocity of $\sim
c/30$.  Equation
(\ref{gravesc}) would predict this total energy for $R\sim 1500$ km. 
In the calculation, $(20\%, 60\%, 90\%)$ of the ejected mass fell
below $(200, 300, 400)$ km before being ejected (C. Fryer, 2000,
personal communication). This indicates that the theory of \S
\ref{S:WD/NS-Theory}, which neglects infall motions and assumes
ejection from a thin outer layer, is not directly comparable to these
AIC models.

Because the total energies are too low and the typical expansion
velocities too slow, \cite{woo92} and \cite{fry99} conclude that AICs
do not cause cosmological GRBs like those listed in Table
\ref{tab:CGRBs}. Although this is a valid conclusion, for completeness we 
investigate the relativistic ejecta from these events, including cases
of more extreme energy release than predicted by \citet{fry99}.

Our $1.38~\Msun$ progenitor model, from \citet{lin99}, uses for its
equation of state a fully degenerate, non-interacting Fermi gas. The
initial density distribution is shown in Figure \ref{fig:denall}. We
note this model has a radius twice that of the WD considered by
\citet{fry99}.

Figure \ref{fig:kewd} shows $\Ek(>\Gamma\beta)$ vs $\Gamma\beta$ for
explosions of different energies in our WD progenitor. As these
simulations neglect gravity and the initial collapse motion of the
progenitor, they should be considered rough approximations to the
result. The lowest curve in this figure matches the mass and energy of
the \citeauthor{fry99} simulations; higher curves address hypothetical
cases with much higher energies. Although an asymmetrical explosion,
which may be expected in the collapse of a rapidly rotating object,
might enhance the energy along the line of sight, we concur with
\cite{woo92} and \cite{fry99} that AICs do not make cosmological
GRBs. We are not in a position to judge whether magnetized winds or
jets from such events \citep{1995ApJ...447..863S} produce GRBs. 

\subsubsection{Neutron Stars}\label{S:NSs}

\begin{figure}
\epsscale{1.0}
\plotone{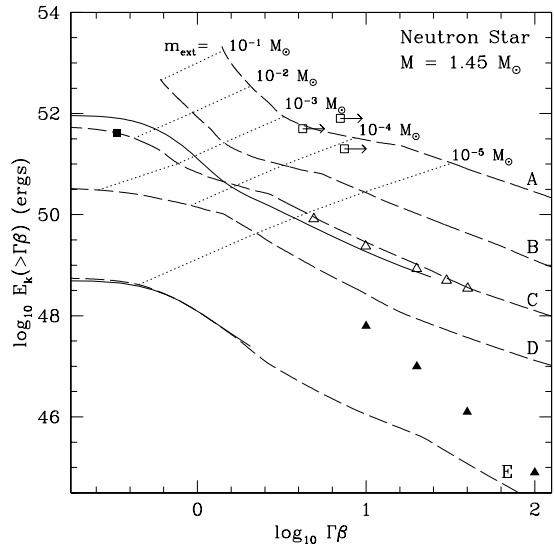}
\caption{
Kinetic energy distribution with velocity for ejecta from neutron star
explosions, after subtracting off the initial gravitational binding
energy. Results for two simulations are shown ({\it solid} lines), and
compared to analytic predictions from \S\ref{S:theory}, integrated
over the progenitor density distribution ({\it long dashed}
lines). The analytic predictions amplify the dependence of velocities
resulting from abrupt changes in the initial density distribution
compared to simulation, which can more accurately handle these
effects. The analytic predictions are used to further explore the
parameter space. Model C, in which ejecta of mass $0.017\:{\rm
M_{\odot}}$ carry away $4\times 10^{51}\:{\rm ergs}$ of kinetic energy
({\it solid square}) is closest to the energetic model of
\citet{fry98}.
The {\it solid triangles} are \citeauthor{fry98}'s predicted upper
limits from nonrelativistic shock propagation models, including a
relativistic correction. Our estimates, including the direct energy
approximation of equation (\ref{Ek-NS-b}) ({\it open triangles}) are
about two orders of magnitude higher. The dotted lines show mass
contours. These reveal the greater ejecta masses that result from the
explosions of higher energy. The {\it open squares} show the
$\gamma-\gamma$ opacity constraint of the three weakest cosmological
GRBs from Table \ref{tab:CGRBs}.
\label{fig:kens}}
\end{figure}

Explosions in neutron stars (NSs) powered by a phase transition to a
``strange'' star have been considered as a potential GRB mechanism
\citep[and references therein]{fry98}. 
\citeauthor{fry98} simulated the collapse and bounce resulting
from a neutron star phase transition to derive properties of the
ejecta. These were then used to implement higher resolution
simulations of shock propagation in the outer 0.01-0.001 ${\rm
M_{\odot}}$, with resolution down to $10^{-11}\:{\rm M_{\odot}}$. This
latter stage made use of a nonrelativistic code, with a simple
approximation based on energy conservation to derive upper limits on
the kinetic energy distribution at high ($>40$) Lorentz
factors. However, this approximation neglects the concentration of
energy in the outermost ejecta at the expense of the inner, due to
pressure gradients in the flow (see \S\ref{shacc}).

\citeauthor{fry98}'s most energetic explosion involved $0.017\:{\rm
M_{\odot}}$ ejected with $\sim 5\times 10^{51}\:{\rm ergs}$ from a
neutron star of initial mass 1.4 ${\rm M_{\odot}}$. Using the Gnatyk
shock formula (eq. \ref{gnatyk}) and their simple estimate of
postshock acceleration,
they inferred an upper limit of $10^{46}\:{\rm ergs}$ for material
traveling with $\Gamma_f>40$, and concluded that NS phase transitions
cannot power cosmological GRBs. However, as these events are
sufficiently energetic to explain some of these bursts, and as the
mean velocity implied by their ejecta is $\sim 0.51 c$ (i.e.,
$\bar\Gamma \simeq 1.16$), we wish to revisit this issue with our
improved treatment of the dynamics.

As the injected energy is small compared to the binding energy of the
NS envelope, and as only a small fraction of this envelope is blown
away, the results of \S \ref{S:WD/NS-Theory} apply. In the simulation
of \cite{fry98}, the stellar surface is shocked after it has fallen
from 11 km to 10 km. We use the low-density Harrison-Wheeler (HW)
equation of state \citep{gle97} to obtain the appropriate value of $n$
underneath $0.017~\Msun$ of material: the pressure at this depth is
$5\times 10^{32}$ dyne cm$^{-2}$, so we adopt the value $n\simeq 12$
appropriate for the highest pressure listed ($10^{31}$ dyne
cm$^{-2}$).  Using these parameters, equation (\ref{gravesc}) predicts
a final kinetic energy of $3.3\times 10^{51}$ erg, sufficiently close
to the value ($\sim 5\times 10^{51}$ erg) quoted by \cite{fry98} to
give us faith in our approximations. As an added check, note that
\citeauthor{fry98} consider a weak explosion that ejects
$10^{-3}\Msun$ with $\sim 3\times 10^{50}$ erg; equation
(\ref{gravesc}) predicts $2.0\times 10^{50}$ erg for this ejecta
mass. Thus, we see that the theory of \S \ref{S:WD/NS-Theory}
underestimates the kinetic energy in the simulations by $\sim
50\%$. This is not due to our approximation of Newtonian gravity, as
\citeauthor{fry98}'s calculations were also Newtonian, but may be due
to our neglect of infall of the outer layers, before they are shocked,
or their neglect of special relativity.

We apply equation (\ref{Ek-NS-b}) to predict the kinetic energy in
ejecta at some higher Lorentz factor. Note a value of $n=4$,
appropriate for the outer distribution, is used in evaluating
$F(\Gamma_f\beta_f)$, while $n=12$, appropriate for the base of the
ejecta, is used to evaluate the preceding terms that normalize the
distribution. Taking $f_{\rm sph}=0.85$, we obtain kinetic energies
of $(8.2, 2.4, 0.85, 0.45, 0.35)\times 10^{49}$ ergs for $\Gamma_f>(5, 10,
20, 30, 40)$; presumably, these energies should be increased by $\sim
50\%$ to account for the discrepancy noted above. This gives an
estimate of $\sim 0.52\times 10^{49}$ erg in material with
$\Gamma_f>40$, more than two orders of magnitude greater than
\citeauthor{fry98}'s inferred upper limit.

We validate this result with direct simulation of relativistic
explosions in a neutron star progenitor, constructed using the
Harrison-Wheeler equation of state at low densities \citep{gle97} and
with the pressure extrapolated to higher densities with a polytrope, using
Newtonian gravity. The central pressure and polytropic index were
varied to reproduce a realistic mass and radius. 
The initial density distribution is shown in Figure
\ref{fig:denall}. Our method follows that employed for explosions in
white dwarfs, except now it is necessary to account for the
progenitor's gravity (\S\ref{S:WD/NS-Theory}). We make a first order
correction to our non-gravitating simulations by subtracting off the
initial gravitational binding energy of ejecta from their final energy
distribution. In Figure \ref{fig:kens} we compare simulation results
with our analytic approximations to the shock dynamics, applied to the
progenitor density distribution, again accounting for the
gravitational binding energy. We also show ({\it open triangles}) the
direct energy approximation (equation \ref{Ek-NS-b}), without the
calibration correction to \citeauthor{fry98} results, which acts to
increase the energies. Our different methods are in good
agreement, and are larger than \citeauthor{fry98}'s estimate ({\it
solid triangles}) by between two and three orders of magnitude, a fact
which highlights the importance of a correct treatment of postshock
acceleration.

For a uniform ambient density of density $n_H \sim 1~{\rm
cm}^{-3}$, equation (\ref{uniformconstraint}) indicates that
$\Gamma_f\simeq \bar\Gamma/2 \gtrsim 44$ if a GRB with
$E_\gamma=10^{52}\:{\rm ergs}$ is to last $\sim 35$ s in its rest
frame, although lower values of $\Gamma_f$ may be acceptable if the
burst occurs in the dense circumstellar environment of a
companion. The photon-photon opacity constraint,
eq. (\ref{photonphoton}), requires $\Gamma_f\simeq
\bar\Gamma/2 \gtrsim 5$ for such a burst, if the photon index is
$\alpha = 2$.  

Considering the cosmological GRBs listed in Table \ref{tab:CGRBs}, we
see that neutron-star phase transitions can match the energies of
several such bursts (970778, 970508, and 000301C) if Lorentz factor
constraints are ignored. However, the required Lorentz factors seem to
be about $\Gamma_f\gtrsim 5-10$ for the slowest part of the ejecta, if
in a sufficiently dense environment. 
The kinetic energy of such ejecta is only 
a few percent of the total, about $10^{50}\:{\rm ergs}$. These
considerations argue against NS phase transitions as the origins of
cosmological GRBs, unless the ejected mass has been significantly
underestimated or the energy can be collimated.

\section{Conclusions}\label{S:conclusions}

Any viable model for the central engines of gamma-ray bursts,
especially for those bursts known to be at cosmological distances,
must explain the existence of large energies ($10^{51-54}$ erg, if
isotropic) in ejecta moving sufficiently fast that, when converted
into radiation, it escapes as non-thermal radiation. Such velocities
require very little ``baryon loading,'' a constraint that many
theories for gamma-ray burst central engines address. In this paper,
we have investigated the possibility that shock hydrodynamical
acceleration, first proposed as a source of GRBs by \cite{col74}, may
concentrate sufficient energy in the fastest ejecta to satisfy this
constraint. We consider ordinary and over-energetic supernova
explosions, making specific models for the hypothesized class of
supernova gamma-ray bursts \citep[S-GRBs;][]{blo98} whose existence
was suggested by SN 1998bw and GRB 980425.  We also calculate the
yield of relativistic ejecta from hypothetical models for
extraordinary ``hypernova'' explosions, as well as possible
accretion-induced collapses of white dwarfs and sudden condensations
of neutron stars.

As an exemplary case, we have considered in detail (\S \ref{S:98bw})
specific models for supernova 1998bw and the gamma-ray burst (GRB
980425) that it may have produced, considering the constraints of the
GRB and the supernova's early radio emission in the context of
progenitor models constructed \citep[by][]{woo99} to fit its light
curve. One goal of this investigation has been to evaluate assertions
\citep{1998ApJ...504L..87W,woo99,hof99} that spherical models for this
explosion cannot explain the GRB. After properly accounting for the
the trans-relativistic shock and postshock dynamics as well as the
interaction with a surrounding stellar wind, we are able to account
for all of the observations of this event with a spherical energetic
Type Ic supernova from a progenitor with high ($\sim {\rm few}\times
10^{-4}\:{\rm M_{\odot}\:yr^{-1}}$) mass loss rate, consistent with
observed carbon-rich (WC) Wolf-Rayet stars.  Our conclusion that the
fastest ejecta need not be asymmetric is reminiscent of the fact that
the radio shell around SN 1993J was spherical
\citep{1997ApJ...486L..31M} despite the assertion
\citep{1996ApJ...459..307H} that its polarization implied an
asymmetrical explosion.
Our model for SN 1998bw and GRB 980425 invokes a circumstellar
environment similar to that inferred around SN 1997cy, which may be
associated with GRB 970514 \citep{2000ApJ...534L..57T}; however the
two-second duration of this burst may be difficult to reconcile with
the external shock hypothesis. 

In order to quantify the process of relativistic mass ejection, we
have extended the nonrelativistic theory of supernova explosions
\citep{mat99} into the relativistic regime. Finding that the
energetics of such ejecta depend sensitively on the velocity of the
shock front as it emerges from the stellar surface, we present in \S
\ref{S:shockprop} a formula for this velocity (eqs. [\ref{relshockf}]
and [\ref{relshock}]) that handles the transition from nonrelativistic
to relativistic flow more accurately than does the prediction of
\citet{gna85} (equation \ref{gnatyk}). Whereas for \cite{mat99} it
sufficed to take a universal value for the normalization of the shock
velocity (and also applied an overall energy conservation constraint
to their ejecta-density models), we have found it necessary to present
(eqs. [\ref{Anr}], [\ref{sigma}], and [\ref{krhoJl}]) a refined
estimate for this coefficient -- one that responds to the initial
density distribution of the progenitor.

The motivation for such precision is the sensitive dependence of the
final velocity of a mass element on the velocity of the shock that
struck it: in the relativistic, planar limit this is $\Gamma_f
\propto \Gamma_s^{2.7}$. This postshock acceleration, neglected in
many works that invoke shock acceleration, is crucial for the
production of appreciable energy in fast ejecta. In \S \ref{shacc} we
present formulae for this postshock acceleration in the planar limit
(eq. [\ref{planar}]) and for the effect of spherical geometry to
reduce the acceleration (eqs. [\ref{fsphdef}] and [\ref{fsph}]). The
latter is somewhat uncertain, due to the limitations of our numerical
calculations; however, since this correction is a factor of order
unity, the energetics of relativistic ejecta are not especially
sensitive to it.
Because we account properly for postshock acceleration, our
predictions for the energetics of relativistic ejecta are much larger
than presented in previous works that considered hydrodynamical shock
acceleration in the context of supernovae, accretion-induced collapses 
of white dwarfs, and phase transitions in neutron stars.

With the dynamics of trans-relativistic mass ejection well in hand, we
turned in \S \ref{S:KEdist} to an evaluation of the amount of kinetic
energy expected in ejecta traveling above some given lower
limit. Equation (\ref{EkGeneral}) 
represents an accurate approximation for this kinetic energy distribution under the
assumption that the outer stellar density distribution is that of a
polytropic atmosphere of index $n$. This formula shows that the
kinetic energy distribution rolls over quite slowly
(fig. \ref{fig:kean}) from $E_k\propto
\beta_f^{-(3.35+5.35/n)}$, to $E_k\propto\Gamma_f^{-(0.58+1.58/n)}$, as
the final velocity in question goes from nonrelativistic
($\Gamma_f\beta_f\ll1$) to relativistic ($\Gamma_f\beta_f\gg1$); this
slow limiting behavior comes from fact that relativistic final
velocities originate from only mildly relativistic shocks. For typical
values of $n$, the kinetic energy in relativistic ejecta decreases as
$\sim 1/\Gamma_f$; this gentle decline with velocity allows for
significant energy at relatively high $\Gamma_f$. 

We next developed (\S \ref{S:whips}) a theory to assess what
characteristics of a stellar envelope make it especially efficient at
converting a given dose of injected energy into relativistic motion --
the astrophysical analog of how to design a good bullwhip. If the
envelope mass is fixed, we found that more centrally-condensed
envelopes are more efficient producers of high-velocity ejecta. If
instead the progenitor star is given, then the energy in fast ejecta
is enhanced by allowing as much of it as possible to collapse
(minimizing $M_{\rm ej}$), as this increases the energy per mass of
what escapes.

Evaluating the energetics of relativistic ejecta for radiative stellar
progenitors in \S \ref{S:whips}, we derived expressions appropriate
for regions dominated by electron scattering (eq. [\ref{ErelThomson}])
or bound-free or free-free (eq. [\ref{ErelKramers}]) absorption,
assuming different possible compositions for the outer envelope. For
the Thomson opacity case, the central concentration of the atmosphere
-- and therefore its efficacy at producing relativistic ejecta -- is
enhanced if the stellar luminosity approaches the Eddington limit.

It should be kept in mind that typical supernovae in supergiant stars
are not capable of creating any relativistic ejecta \citep{mat99},
because there, shock acceleration ends at nonrelativistic velocities
at a depth that matches the width of the shock. In this paper, we have
only considered progenitors sufficiently compact and exploding with
enough energy, that relativistic ejection does occur.

Although we do not model asymmetrical explosions, our theory for shock
and postshock dynamics can be applied to simulations of such an
event. In \S \ref{S:aspherical} we present a formula
(eq. [\ref{eq:asphYp}]) for the relativistic yield of each patch of
the stellar surface, in terms of the intensity of the
(nonrelativistic) shock as it might be observed in such a
simulation. As multidimensional simulations cannot yet afford to treat
relativistic dynamics with the extremely fine zoning that would be
required to match the predictive power of this formula, it could
be quite useful in future investigations.

In \S \ref{S:WD/NS-Theory} we have put our dynamical formulae to
another use: the prediction of what will be ejected in the case of an
explosion so weak that it only ejects a small mass from the outside of
the star. Assuming ballistic motion after a brief phase of
acceleration, the escape velocity marks the inner boundary of the
ejecta. For this reason, the overall kinetic energy
(eq. [\ref{gravesc}]) and the kinetic energy above some given velocity
(eqs. [\ref{Ek-WD/NS}], [\ref{Ek-WD-a}], and [\ref{Ek-NS-b}]) are
proportional the initial binding energy of the ejecta. Applied to
neutron star phase transitions in \S \ref{S:NSs}, equation
(\ref{gravesc}) predicts within $50\%$ the average kinetic energy per
gram in the ejecta in \cite{fry98}'s simulations. These considerations
also provide some insight into the characteristic shock velocity at
the base of the ejecta (\S \ref{S:CharacteristicVelocities}).

After calculating in \S \ref{SS:98bw} the energy distribution of
relativistic ejecta from \cite{woo99}'s model CO6 for SN~1998bw, and
demonstrating that this can explain both the observed gamma-ray burst
($\sim 35$ s timescale; \S \ref{S:gammarays}) and the rapidly
expanding radio shell ($\sim 12$ day timescale; \S \ref{S:radio}), we
turned to the modeling of far more energetic ``hypernova'' explosions
in \S \ref{S:C-GRBs}. The association of GRB afterglows with the
stellar population of their host galaxies \citep{blo00b}, and more
tentatively with star forming regions \citep{kul00, djo00, gal00b},
argues in favor of a connection with the deaths of massive stars and
against scenarios involving neutron star mergers. Furthermore, the
difficulty of avoiding baryon loading while trying to escape the
stellar progenitor, encountered even in highly asymmetric jet models
\citep{alo00}, is overcome in the hydrodynamic shock acceleration
mechanism.

In this model GRBs, like their afterglows, would represent radiation
from an external shock that accelerates the ambient medium, with an
extra component (of comparable luminosity) from the reverse shock that
decelerates the ejecta; both shocks must decelerate
monotonically. Note that the higher efficiency of $\gamma$-ray
production in models including external shocks, compared to those
based solely on internal shocks \citep{kum99}, gives external shock
models less stringent total energy requirements.  The spectral
signature of such two-shock structures \citep{2000ApJ...535L..33S}
should be used to test this hypothesis; however, the reverse shock is
buried under more material (by a factor of $\Gamma$) and may not be
visible early on.  While highly structured GRB profiles are difficult
to accommodate in this scenario \citep{sar97}, most of the GRBs with
afterglows have relatively smooth pulses. Small amounts of
substructure may result from interaction with an inhomogeneous
medium \citep{der99,fen99}, or from postshock velocity perturbations
produced by instabilities in the accelerating shock
\citep{1994ApJ...435..815L} or the presence of steep density
shelves in the progenitor's atmosphere. More detailed modeling of
these processes is required. In some bursts, there is evidence that a
smooth and soft component of the GRB flux is an early manifestation of
the afterglow \citep{1999A&AS..138..443B, 1999ApJ...524L..47G,
2000AstL...26..269B, 2000A&A...358L..41T}. The existence of this
component agrees with the external-shock model, although the fact that
it does not account for the entire GRB does not. GRB 990123 had an
optical flash that has been interpreted
\citep[e.g.,][]{2000ApJ...535L..33S} to arise from reverse-shock
emission; however, this model invoked a brief rather than a continuous
collision of ejecta with the swept-up shell.

Because an external medium is necessary to liberate the energy, the
Lorentz factor of the emitting region must not only exceed the usual
lower limit from photon-photon opacity (eq. [\ref{photonphoton}]), but
also a limit due to the opacity of a circumstellar wind
(eq.~[\ref{GammaOpticallyThin2}]) or the density of a uniform
background (eq. [\ref{uniformconstraint}]). The observed gamma-ray
energies thus become, in a spherical model for the burst, lower limits
for the amount of energy in ejecta whose Lorentz factor satisfies
these constraints. As the bursts in question have quite smooth time
profiles, and because our adoption of the external-shock hypothesis
dictates that the age of a burst, rather than the timescale of its
substructure, be used to constrain the Lorentz factor, we predict
smaller minimum Lorentz factors than commonly quoted in the
literature, by factors of ten or more. This eases the baryon-loading
constraint by a similar amount. 

We find that a simple, and admittedly extreme, spherical hypernova
model satisfies the energy and velocity constraints of many of the
observed cosmological GRBs. The most energetic bursts, such as GRB
990123, require asymmetric explosions. These need not be highly
aspherical as the results of \S
\ref{S:aspherical} show that a mild (factor of two) variation in shock 
intensity at a given column below the surface of the star can lead to
an extreme (factor of 100) variation in the observed burst
intensity. If such variations are justified by simulations, this
result may allow even the most energetic bursts to be explained by
hydrodynamical shock acceleration in aspherical hypernovae.  Since the
very luminous bursts do not differ from typical ones in any
significant manner other than luminosity, it is likely that all
cosmological bursts are aspherical, and as a result there is no need
to invoke hypernova models as extreme as the one we have considered.

Since we predict a kinetic energy that decreases with Lorentz factor
roughly as $1/\Gamma_f$, there must be more energy in slower
relativistic ejecta, and (if $\Ein\ll M_{\rm ej}c^2$) much more in
nonrelativistic ejecta as well. As the shocked region decelerates, an
increasing fraction of this energy should appear in the afterglow;
therefore, estimates of afterglow energetics \citep[e.g.,][]{fre99}
may potentially rule out such models in the future. However, having
demonstrated in equation (\ref{EshellWindShock}) that the energy of
the shocked shell increases quite slowly, we find that
\citeauthor{fre99}'s constraints cannot yet rule out
hypernovae as sources of cosmological GRBs.

Energetic events in compact objects have also been considered as
engines for GRBs. White dwarfs (\S \ref{S:WDs}) explode as type Ia
supernovae, but the relativistic ejecta from SNe Ia are too weak to be
of much interest. Accretion-induced collapse of white dwarfs can
potentially concentrate a somewhat smaller energy ($\sim 10^{50}$ erg)
in a smaller ($\sim 0.1 ~\Msun$) mass of ejecta. Considering the
simulations of \cite{fry99}, we predict a much greater yield of
relativistic ejecta than their estimates; but not enough to power
cosmological GRBs.

Neutron star phase transitions, considered in \S \ref{S:NSs}, are well
described by the weak explosion theory of \S \ref{S:WD/NS-Theory}. As
equation (\ref{gravesc}) gives the overall energetics of the explosion
to within $50\%$, we are justified in using equation (\ref{Ek-WD/NS})
to predict the energetics of fast ejecta. Our estimates, validated
with numerical simulation, are several orders of magnitude higher than
those of \cite{fry98}, who underestimated postshock acceleration. However,
we note that such events do not put enough energy at sufficient
Lorentz factor to satisfy the observational constraints for distant
bursts with afterglows.

\acknowledgements We thank Stan Woosley and Lee Lindblom for kindly
providing models of SN 1998bw's progenitor star and a white dwarf,
respectively. We also thank M. Aloy, A. Filippenko, Z.-Y. Li and the
referee for detailed comments on the manuscript. We are grateful to
A. Cumming, R. Fisher, P. Kumar, A. MacFadyen, R. Sari,
E. Scannapieco, N. Shaviv, and A. Youdin for helpful discussions. CDM
appreciates the gracious hospitality of Sterl Phinney and Roger
Blandford during his visits to Caltech, as well as financial support
from NSERC. The research of JCT and CFM is supported by NSF grant AST
95-30480.

\end{document}